\documentclass[prd,twocolumn,floatfix,preprintnumbers,reprint,showpacs]{revtex4}
\usepackage{epsfig,amsmath,amssymb,bm,color,graphicx}
\usepackage{afterpage}
\usepackage{pdfpages}

\usepackage{newtxtext,newtxmath}
\usepackage[caption=false]{subfig} 
\usepackage[T1]{fontenc}
\usepackage[normalem]{ulem}
\usepackage{lastpage}
\usepackage{soul}

\def\lya{Lyman-$\alpha$\,}

\newcommand{\Mpch}{h^{-1}\,\mathrm{Mpc}}
\newcommand{\skm}{\mathrm{km^{-1}\,s}}
\newcommand{\kms}{\mathrm{km\,s^{-1}}}

\newcommand{\ditto}{{\tt "}}

\newcommand{\be}{\begin{equation}}
\newcommand{\ee}{\end{equation}}

\usepackage{xcolor}

\usepackage[hidelinks]{hyperref}
\def\orcid#1{\href{https://orcid.org/#1}{\includegraphics[keepaspectratio,width=0.7em]{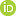}}}

\newcommand{\orcidauthorA}{\orcid{0000-0002-5445-461X}} 
\newcommand{\orcidauthorB}{\orcid{0000-0002-2642-5707}} 
\newcommand{\orcidauthorC}{\orcid{0000-0001-8443-2393}} 
\newcommand{\orcidauthorD}{\orcid{0000-0003-2764-8248}} 
\newcommand{\orcidauthorE}{\orcid{0000-0002-1841-4274}} 
\newcommand{\orcidauthorF}{\orcid{0000-0001-8778-7587}} 
\newcommand{\orcidauthorG}{\orcid{0000-0002-8340-6537}} 
\newcommand{\orcidauthorH}{\orcid{0000-0003-2344-263X}} 
\newcommand{\orcidauthorI}{\orcid{0000-0002-2423-7905}} 
\newcommand{\orcidauthorJ}{\orcid{0000-0001-5211-1958}} 
\newcommand{\orcidauthorK}{\orcid{0000-0001-5829-4716}} 

\graphicspath{{figures/}}

\begin{document}

\title{Unveiling Dark Matter free-streaming at the smallest scales with high redshift Lyman-alpha forest}
\author{Vid Ir\v{s}i\v{c}$^{1,2}$\,\orcidauthorA,\thanks{E-mail: vi223@cam.ac.uk (VI)}
Matteo Viel$^{3,4,5,6,7}$\,\orcidauthorB,
Martin G. Haehnelt$^{1,8}$\,\orcidauthorC, 
James S. Bolton$^{9}$\,\orcidauthorD,
Margherita Molaro$^{9}$\,\orcidauthorE,
Ewald Puchwein$^{10}$\,\orcidauthorF,
Elisa Boera$^{5,6}$\,\orcidauthorG,
George D. Becker$^{11}$\,\orcidauthorH,
Prakash Gaikwad$^{12}$\,\orcidauthorI,
Laura C. Keating$^{13}$\,\orcidauthorJ,
Girish Kulkarni$^{14}$\,\orcidauthorK
}

\smallskip
\affiliation{
$^{1}$Kavli Institute for Cosmology, University of Cambridge, Madingley Road, Cambridge CB3 0HA, UK\\
$^{2}$Cavendish Laboratory, University of Cambridge, 19 J. J. Thomson Ave., Cambridge CB3 0HE, UK\\
$^{3}$SISSA- International School for Advanced Studies, Via Bonomea 265, 34136 Trieste, Italy\\
$^{4}$INFN – National Institute for Nuclear Physics, Via Valerio 2, I-34127 Trieste, Italy\\
$^{5}$IFPU, Institute for Fundamental Physics of the Universe, via Beirut 2, 34151 Trieste, Italy\\
$^{6}$INAF, Osservatorio Astronomico di Trieste, Via G. B. Tiepolo 11, I-34131 Trieste, Italy\\
$^{7}$ICSC - Centro Nazionale di Ricerca in High Performance Computing, Big Data e Quantum Computing, Via Magnanelli 2, Bologna, Italy\\
$^{8}$Institute of Astronomy, University of Cambridge, Madingley Road, Cambridge CB3 0HA, UK\\
$^{9}$School of Physics and Astronomy, University of Nottingham, University Park, Nottingham, NG7 2RD, UK\\
$^{10}$Leibniz-Institut f\"ur Astrophysik Potsdam, An der Sternwarte 16, 14482 Potsdam, Germany\\
$^{11}$Department of Physics and Astronomy, University of California, Riverside, CA 92521, USA\\
$^{12}$Max-Planck-Institut für Astronomie, Königstuhl 17, D-69117 Heidelberg, Germany\\
$^{13}$Institute for Astronomy, University of Edinburgh, Blackford Hill, Edinburgh, EH9 3HJ, UK\\
$^{14}$Tata Institute of Fundamental Research, Homi Bhabha Road, Mumbai 400005, India\\
}

\begin{abstract}
This study introduces novel constraints on the free-streaming of thermal relic warm dark matter (WDM) from \lya{} forest flux power spectra. Our analysis utilises a high-resolution, high-redshift sample of quasar spectra observed using the HIRES and UVES spectrographs ($z=4.2-5.0$). We employ a Bayesian inference framework and a simulation-based likelihood that encompasses various parameters including the free-streaming of dark matter, cosmological parameters, the thermal history of the intergalactic medium, and inhomogeneous reionization, to establish lower limits on the mass of a thermal relic WDM particle of $5.7\;\mathrm{keV}$ (at 95\% C.L.). This result surpasses previous limits from the \lya{} forest through reduction of the measured uncertainties due to a larger statistical sample and by measuring clustering to smaller scales ($k_{\rm max}=0.2\;\skm$). The approximately two-fold improvement due to the expanded statistical sample suggests that the effectiveness of \lya{} forest constraints on WDM models at high redshifts are limited by the availability of high-quality quasar spectra. Restricting the analysis to comparable scales and thermal history priors as in prior studies ($k_{\rm max}<0.1\;\skm$) lowers the bound on the WDM mass to $4.1\;\mathrm{keV}$. As the precision of the measurements increases, it becomes crucial to examine the instrumental and modelling systematics. On the modelling front, we argue that the impact of the thermal history uncertainty on the WDM particle mass constraint has diminished due to improved independent observations.  At the smallest scales, the primary source of modeling systematic arises from the structure in the peculiar velocity of the intergalactic medium and inhomogeneous reionization.
\end{abstract}

\maketitle
\section{Introduction}
The \lya forest is the main manifestation of the high-redshift intergalactic cosmic-web. It is visible in the spectra of quasars (QSOs) and produced by the scattering of the background photons with the neutral hydrogen atoms along the line-of-sight \citep{meiksin09,mcquinn16}.
The \lya forest is a unique probe of geometry 
and the dynamical state of the Universe, probing diffuse matter around galaxies and in the intergalactic medium (IGM) in regimes which
are not covered by other observables, both in terms
of redshifts and scales.

In the last decade we have witnessed tremendous progress in the cosmological investigation of the \lya forest, mainly along two different directions which are connected to fundamental physics.
For example, the discovery of Baryonic Acoustic Oscillations in the 3D correlation function of the transmitted flux has offered the possibility to constrain new physics beyond the standard cosmological model, in the context of allowing curvature or an evolution of the equation of state for dark energy \citep{aubourg}.
Another important research line, following the work of \citep{nara2000,seljak06}, has focussed on the 1D flux power spectrum used to probe the growth of structure down to the smallest scales to see to which extent dark matter free streaming could be constrained.

In this work we will investigate this second aspect and present new results based on a new set of simulations which incorporate 
the most important physical ingredients \citep{Puchwein2023,Molaro2022}, and a new comprehensive analysis of high-resolution high-redshift data down to the smallest scales.
A key goal is to disentangle the different roles of the physical processes able to affect the 1D flux power: the thermal broadening, which is a 1D effect acting along the line-of-sight and is sensitive to the instantaneous gas temperature, and two 3D effects, the gas pressure smoothing that depends on the whole thermal history of the IGM and the dark matter (DM) free streaming.

The possibility of constraining the nature of DM by using the \lya forest has motivated a series of works which were able to constrain the models further, explore different particle physics dark matter candidates, and combine likelihoods with other experiments able to constrain the nature of dark matter with strong lensing or flux ratio anomalies \citep{enzi21,hsueh}.
One of the main reasons to explore warm dark matter (WDM) models was to solve or ease putative problems of cold dark matter at small scales \citep{colin00,bode01}. However most of these tensions must be discussed also in the context of baryonic physics \citep{wdmreview}, with processes like galactic feedback playing a major role.
Moreover, it appears that minimal extensions of the standard model of particle physics could also accommodate particles like sterile neutrinos or a scalar field \citep{whitepapersterile,snowmass22,Rogers2021,Irsic2017c}, which could suppress or erase power at small scales, effectively acting as WDM.

For thermal WDM masses in the keV range, the power suppression happens at the small non-linear scales sampled by the \lya forest. In particular, QSO data sets with different resolution and signal-to-noise properties have been used in order to tighten the constraints. The low resolution SDSS and BOSS data sets \citep{seljak06,baur17,palanque20}, the medium resolution X-Shooter sample \citep{Irsic2017a} and the high resolution and high signal-to-noise Keck/HIRES and UVES/VLT QSO spectra \citep{Viel05,Viel13wdm,Irsic2017b,Garzilli2021,Villasenor2022b} have all played a major role in the advancement of the field.
For example, while the low and medium resolution data are not particularly effective in sampling the scales of the cutoff fully, they nevertheless are sensitive to the thermal history and can return very tight constraints especially when combined with data at smaller scales.
The goal of this paper is to give a comprehensive state-of-the-art analysis focusing on high resolution data \cite{Boera2019}.
In Section \ref{sec:data} we will describe the data set, while in Section \ref{sec:simulations} we will present the suite of hydrodynamical simulations used. Section \ref{sec:results} will contain our new results which will be extensively discussed in terms of the thermal history of the IGM, the dependence on mass resolution, patchy reionization, instrumental effects (including modelling of the noise) and consistency with results in the literature. We will conclude in section \ref{sec:conclusions}.

\section{Data}
\label{sec:data}
We apply our analysis to the measurements presented in \citep{Boera2019}. Their 1D flux power spectrum is estimated using 15 high signal-to-noise spectra observed by VLT/UVES \citep{uves} and Keck/HIRES \citep{hires}. The measurements span the high redshift range of $z=4.2 - 5.0$ in bins of $\Delta z = 0.4$. In each of the redshift bins the flux power spectrum is measured in 15 $k$-bins equidistantly spaced in $\log_{10}k$ in the range of $\log_{10}\left(k/[\skm]\right) = -2.2$ to $\log_{10}\left(k/[\skm]\right) = -0.7$, with logarithmic spacing of $\Delta \log_{10}\left(k/[\skm]\right) = 0.1$. Unless specified otherwise we use the full extent of the data, resulting in $45$ data points across three redshift bins. 

The spectrograph resolution in these observations is very high, with $R\sim50,000$ (FWHM of $\sim 6\,\kms$ for HIRES and $\sim 7\,\kms$ for UVES). As already pointed out in the study of \citep{Boera2019} the effects of resolution uncertainty are very small, even for the highest wavenumber power spectrum bin measured. A conservative estimate of the 10\% uncertainty on the resolution leads only to 1\% (5\%) uncertainty on the 1D flux power spectrum at scales of $k = 0.1\,(0.2)\,\skm$. The power spectrum measurements of \citep{Boera2019} were reported both with and without instrumental resolution correction. In this analysis we use the measurements with instrumental resolution corrected, and propagate this correction through the covariance matrix. The reported measurements are also corrected for power spectrum due to metal contaminants.

The typical flux noise estimated in these measurements is white noise, with its power spectrum amplitude of $0.1 - 0.2\,\kms$. This is comparable to the estimated level of the models at the highest wavenumbers. Characterizing and accounting for the noise levels is of key importance, and has been one of the factors restricting previous analyses to smaller wavenumbers.

The $k$ range of \citep{Boera2019} covers the smallest scales measured with the 1D flux power spectrum, extending to $k\sim0.2\,\skm$, a factor of two higher wavenumber than in previous studies \citep{Viel13wdm,Irsic2017b,Garzilli2021}. These studies have shown that the constraining power on WDM models from the Ly$\alpha$ forest is dominated by high redshifts and the smallest scales, making this an ideal data set to exploit.

\section{Simulations}
\label{sec:simulations}

\begin{table*}
\centering
\caption{List of simulations used in this work (see also \citep{Puchwein2023}). From left to right, the columns list the simulation name, the box size in $h^{-1}\rm\,cMpc$, the number of particles, the redshift of reionisation (defined as the redshift when the volume averaged ionised fraction $1 - x_{\rm HI} \leq 10^{-3}$), the gas temperature at the mean density, $T_{0}$, the cumulative energy input per proton mass at the mean density, $u_{0}$, for $4.6\leq z \leq 13$ \citep[cf.][]{Boera2019}, and the cosmological model described by $\Lambda$CDM parameters ($\sigma_8$,$n_s$) and a WDM parameter for the inverse of the WDM particle mass of a thermal relic ($m_{\rm WDM}^{-1}$). The upper section of the table lists the models in the first set of simulations that we use for our MCMC analysis (see text for details). The lower section of the table lists our second set of simulations, which includes mass resolution (R10) and box size (B40) corrections to the predicted flux power spectrum. The dark matter and gas particle mass are $5.37\times10^{5}$ $h^{-1}\,M_{\odot}$ and $9.97\times10^{4}$ $h^{-1}\,M_{\odot}$ respectively for L20, B40 and a subset of R10 runs (2$\times 512^3$). The cosmology parameter ranges for $\sigma_8$ include five runs $[0.754,0.804,0.829,0.854,0.904]$ and similarly 5x runs for $n_s$ $[0.921,0.941,0.961,0.981,1.001]$. The WDM mass in $\mathrm{keV}^{-1}$ of $0$ indicates a CDM run. The other WDM runs are for $2,3$ and $4$ $\mathrm{keV}$ WDM particle mass.
}
\label{table:summary_sims}
\begin{tabular}{lcccccccc}
\hline
Name     & $L_{\rm box}$ & $N_{\rm part}$ & $z_{\rm rei}^{\rm end}$ & $T_{0}(z=4.6)$ & $u_{0}(z=4.6)$ & $\sigma_8$ & $n_s$ & WDM mass\\
 & $[h^{-1}\rm\,cMpc]$ & &  & [$\rm K$] &  $[\rm eV\,m_{\rm p}^{-1}]$ & & & $[\mathrm{keV}^{-1}]$  \\
\hline 
L20-ref      & 20.0 & $2\times 1024^{3}$ & 6.00 & 10066 & 7.7 & $[0.754-0.904]$ & $[0.921-1.001]$ & $[0,\frac{1}{4},\frac{1}{3},\frac{1}{2}]$ \\
L20-late & \ditto & \ditto &  5.37 & 10069 & 6.6 & \ditto & \ditto & \ditto \\
L20-early & \ditto & \ditto &  6.70 & 10050 & 9.6 & \ditto & \ditto & \ditto\\
L20-very early & \ditto & \ditto &  7.40 & 10003 & 11.4 & \ditto & \ditto & \ditto\\
L20-ref-cold & \ditto & \ditto &  5.98 & 6598 & 4.3 & \ditto & \ditto & \ditto\\
L20-late-cold & \ditto & \ditto &  5.35 & 6409 & 3.6 & \ditto & \ditto & \ditto\\
L20-early-cold & \ditto & \ditto &  6.69 & 6803 & 5.4 & \ditto & \ditto & \ditto\\
L20-very early-cold & \ditto & \ditto &  7.39 & 6806 & 6.4 & \ditto & \ditto & \ditto\\
L20-ref-hot & \ditto & \ditto &   6.01 & 13957 &14.4 & \ditto & \ditto & \ditto\\
L20-late-hot & \ditto & \ditto & 5.38 & 13451 &12.5 & \ditto & \ditto & \ditto\\
L20-early-hot & \ditto & \ditto &  6.71 & 14369 & 17.8 & \ditto & \ditto & \ditto\\
L20-very early-hot & \ditto & \ditto &  7.41 & 14624 & 21.1 & \ditto & \ditto & \ditto\\
\hline
B40-ref  & 40.0 & $2\times 2048^{3}$ &  6.00 & 10063 & 7.7 & $0.829$ &  $0.961$ & 0 \\
\hline
R-set & [5.0,10.0,20.0] & $2\times$[$1024^{3}$,$768^3$,$512^3$] & 6.00 & 10066 & 7.7 & $0.829$ & $0.961$ & $0$ \\ 
R10-ref      & 10.0 & $2\times$[$1024^{3}$,$512^3$] & 6.00 & 10066 & 7.7 & $0.829$ & $0.961$ & $[0,\frac{1}{4},\frac{1}{3},\frac{1}{2}]$ \\
R10-late & \ditto & \ditto &  5.37 & 10069 & 6.6 & \ditto & \ditto & \ditto \\
R10-early & \ditto & \ditto &  6.70 & 10050 & 9.6 & \ditto & \ditto & \ditto\\
R10-ref-cold & \ditto & \ditto &  5.98 & 6598 & 4.3 & \ditto & \ditto & \ditto\\
R10-late-cold & \ditto & \ditto &  5.35 & 6409 & 3.6 & \ditto & \ditto & \ditto\\
R10-ref-hot & \ditto & \ditto &   6.01 & 13957 &14.4 & \ditto & \ditto & \ditto\\
\hline
\end{tabular}
\end{table*}

The absorption features of the Ly$\alpha$ forest contain a wealth of information regarding cosmology and the nature of dark matter, as well as the thermal state of the intergalactic gas. Due to the high sensitivity of the spectrographic instruments it provides a unique window into clustering at the smallest scales. Accessing that information, however, is a non-trivial task. The standard approaches of clustering analysis that invoke biasing schemes typically rely on perturbation theory \citep{Garny20} or build an approximate clustering scheme \citep{Irsic18}. While very informative in a qualitative sense, these schemes cannot capture the complexity of the data that is highly sensitive to non-linear structure evolution and gas physics, such as Doppler broadening and thermal pressure smoothing \citep{Gnedin98}.

Such a task requires simulating the expected Ly$\alpha$ forest in different thermal and cosmological models, spanning a wide, multi-dimensional parameter space, and comparing it to the data. In this work we carry out the comparison within the framework of Bayesian inference analysis, which describes -- according to Bayes' theorem -- the posterior probability $p(\theta|D)$ having parameters $\theta$ given observed data $D$ as:
\begin{equation}
    p(\theta|D) \propto {\cal L}(D|\theta) \times \pi (\theta) \,,
\end{equation}
where ${\cal L}(D|\theta)$ is the likelihood and $\pi(\theta)$ is the prior on each parameter.

In this work we expand upon the Bayesian inference set-up adopted in \citep{Molaro2022,Molaro2023} to evaluate the likelihood and prior at each parameter combination in the sampler. The likelihood is evaluated jointly at all the observed data points. This is based on the Monte Carlo Markov Chain (MCMC) sampler, combined with the Metropolis-Hastings algorithm by dynamically learning the proposal matrix from the covariance that was introduced in \citep{Irsic2017b}. The precision of the thermal parameter recovery with this simulation based emulator was shown to be in good agreement with more advanced machine-learning augmented emulator models \citep{Molaro2023}.

The priors $\pi(\theta)$ we adopt in our analysis are described in section ~\ref{sec:results}. The likelihood is modelled as a Gaussian likelihood, determined by the data and its covariance, and a theoretical prediction for the flux power spectrum. The latter is estimated using hydro-dynamical numerical simulations.

We use simulations from the Sherwood-Relics project \citep{Puchwein2023}. These are a series of high-resolution cosmological hydro-dynamical simulations that use a customized version of \texttt{P-Gadget3} (see \citep{springel05} for the original \texttt{Gadget-2} reference). We use cosmological boxes of size 20 $\Mpch$ with $2 \times 1024^3$ dark matter and gas particles. The box size and resolution have been chosen to adequately resolve the small scale structure that contributes to the flux power spectrum of the Ly$\alpha$ forest, while still retaining a cosmologically relevant volume \citep{bolton09,lukic15,bolton17,Doughty2023}. We further correct the numerical convergence with both box size and resolution with a series of additional simulations summarized in Table~\ref{table:summary_sims}. In all models we use a simple, computationally efficient star-formation scheme -- often called \texttt{Quick\_lya} -- where gas particles are converted into collisionless star particles if they reach overdensities $\Delta = 1 + \delta > 10^3$ and temperatures $T < 10^5\;\mathrm{K}$ \citep{Viel04}. We assume a flat $\Lambda$CDM cosmology with $\Omega_\Lambda=0.692$, $\Omega_m=0.308$, $\Omega_b = 0.0482$, $\sigma_8=0.829$, $n_s=0.961$, $h=0.678$, and a primordial helium mass abundance of $Y_p = 0.24$ \citep{planck18}. The initial conditions for the CDM simulations are identical to those used in the earlier Sherwood simulation project \citep{bolton17}. We use the WDM transfer function approximation of \citep{Viel13wdm}.

A set of simulations is constructed using modifications to the spatially uniform UV background synthesis model introduced by \citep{Puchwein19}. These simulations are similar to models used in earlier works \citep{Viel13,Nasir16,Irsic2017b,Irsic2017c}, with the main improvements being the larger dynamic range of the simulations, the use of a non-equilibrium thermo-chemistry solver \citep{Puchwein15}, and improved treatment of the IGM opacity that consistently captures the transition between neutral and ionised IGM.

In addition to running a model with the fiducial UV background, we also vary the photo-heating rates to achieve models with different gas temperatures and ends of reionization, following the approach described in \citep{Becker11,Puchwein2023}. This approach results in 12 models with varying thermal histories (see Table.~\ref{table:summary_sims}). For {\it each} of the thermal history models with fiducial $\Lambda$CDM cosmology, we also run models varying the WDM particle mass ($m_{\rm WDM}=[2,3,4]\;\mathrm{keV}$), amplitude of $\Lambda$CDM matter clustering ($\sigma_8=[0.754,0.804,0.854,0.904]$) and spectral index of inflation ($n_s=[0.921,0.941,0.981,1.001]$). This results in a total of $12 \times (3 + 2\times4 + 1) = 144$ simulations.

In order to construct a sufficiently well sampled grid of models spanning the entire multi-dimensional parameter range, we post-process the 144 simulations (12 simulations for each cosmology) to obtain different parameter combinations. We follow the method of \citep{Boera2019,Gaikwad20} in order to interpolate in the temperature-density plane. Briefly, we rotate and translate the line-of-sight particles in the temperature-density plane to obtain models with different temperature at mean density $T_0$ and temperature-density power-law indices $\gamma$ (the values of $T_0$ and $\gamma$ are inferred from the line-of-sight gas properties; a power-law relation is fitted to points in the temperature density plane in tha range of gas overdensity ($0.1<\Delta_g<1.0$) and neutral fraction weighted gas temperature ($T<10^5\;\mathrm{K}$)). This preserves the temperature-density cross-correlation coefficient, allowing for an inexpensive construction of models with different thermal parameters on a finely spaced grid. 

In post-processing we also vary the redshift evolution of the mean transmission $\langle F\rangle$, by rescaling the optical depth of Ly$\alpha$ absorption ($\tau_{\rm Ly\alpha}$) obtained from simulations to match observed values of the effective optical depth $\tau_{\rm eff} = -\ln\langle F\rangle$. Uncertainties in the background photo-ionization rate mean a rescaling is commonly used to match the simulations to observations \citep{Bolton05,lukic15}. Note that this step is only a good approximation after reionization, as it implicitly assumes that the gas in the low density IGM is in photo-ionization equilibrium, such that $\tau_{\rm Ly\alpha} \propto x_{\rm HI} \propto \Gamma_{\rm HI}^{-1}$. The redshift evolution that we adopt for $\tau_{\rm eff}$ is:
\begin{equation}
    \tau_{\rm eff} = 1.56 \times \left(\frac{1+z}{5.75}\right)^4,
    \label{eq:teff}
\end{equation}
taken from \citep{Boera2019}, and similar to the evolution reported in \citep{Viel13,Becker13}.

Using the methods described above, we construct a $15\times10\times10$ grid of parameter values on top of {\it each} of the 144 simulations (upper section of Table.~\ref{table:summary_sims}). This grid of models consists of 10 values of $T_0$ spanning the range from 5,000 to 15,000 K in steps of 1,000 K; 10 values of $\gamma$ spanning the range from 0.9 to 1.8 in steps of 0.1; and 15 values of $\tau_{\rm eff}$ in the range from 0.3 to 1.8 times the value in Eq.~\ref{eq:teff}, in multiplicative steps of 0.1. This gives a total of $15\times10\times10\times12\times(1+3+2\times4) = $216,000 models. Since we do not extrapolate outside of this grid of models, we have implicit priors on $T_0$ between 5,000 and 15,000$\;\mathrm{K}$, $\gamma$ between 0.9 and 1.8, and for $u_0$ between $(4.03,21.12)$, $(3.65,21.08)$ and $(2.46,18.73)\;\mathrm{eV/m_p}$ for redshifts $4.2$, $4.6$ and $5.0$, respectively.

\subsection{Flux power spectrum models}

\begin{figure*}
    \centering
    \captionsetup{justification=centering}
    \includegraphics[scale=0.2]{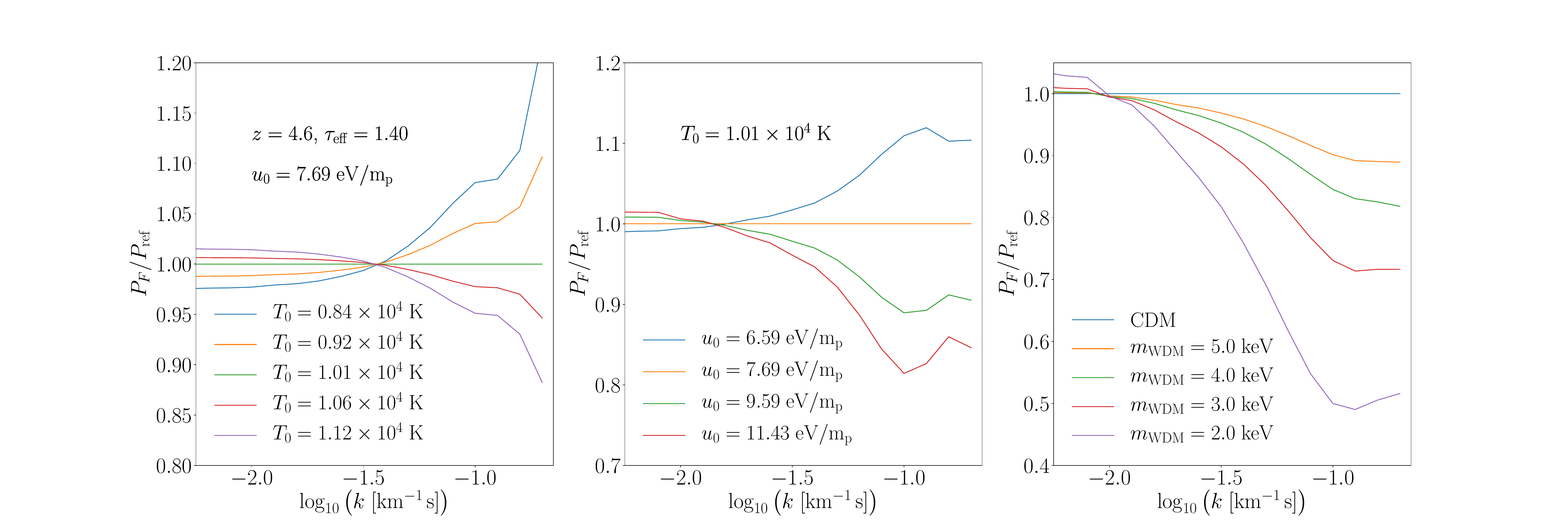}
    \captionsetup{justification=justified}
    \caption{The relative ratios of the 1D flux power spectra of the simulated models relative to a reference simulation run when varying one parameter at a time: $T_0$ (left), $u_0$ (center) and $m_{\rm WDM}$ (right). In each panel all the other parameters are kept fixed. The scale-dependence of the flux power spectrum changes in response to changes in the input parameters. The left panel shows the effect of thermal broadening on the absorption features of the forest. The center and right panels show the emergence of a small-scale enhancement of the relative flux power spectrum in simulations with varying reionization history and WDM free-streaming. The cumulative heat injection values (center panel) correspond to reionization ending at $5.25,6.0,6.75$ and $7.5$ (top to bottom) for the ionizing UV background model of \citep{Puchwein19}. 
    }   
    \label{fig:p1d_models_2}
\end{figure*}

From the grid of models we extract 5,000 lines of sight in different orientations through the box. The flux field along each skewer is Fourier transformed, and the resulting power spectrum is averaged over all the lines of sight, resulting in the predicted 1D flux power spectrum for a given model. In order to compare the simulated models to the data we construct an emulator that interpolates the 1D flux power spectrum between the models, allowing us to explore the parameter space spanned by the simulated models. The emulator is based on linear interpolation \citep{Irsic2017b}. Since the grid of models fills the parameter space in a uniform fashion the interpolation error is small as demonstrated on the sub-set of the models in \citep{Molaro2022}. Neglecting quadratic terms in the interpolation leads to at most 1.2\% correction at high $(k,z)$ in the flux power spectrum, well below the statistical uncertainty on the data.

Fig.~\ref{fig:p1d_models_2} shows the 1D flux power spectra when varying the parameters that govern the three main scales of suppression of the flux power. In the left panel, increasing the temperature of the gas at mean density increases the suppression on small scales (high-$k$), while inducing a small increase in power at large scales (low-k). The latter is due to keeping $\tau_{\rm eff}$ fixed, while the former can be understood in the context of thermal broadening of the lines -- the transmission profile of the \lya{} scattering is determined by the random motion of the gas at a finite temperature. The higher the temperature the larger the velocity dispersion of the thermal motion, leading to more extended profiles that erase small-scale structure.

A related effect, shown in the central panel of Fig.~\ref{fig:p1d_models_2}, is the effect of pressure smoothing. As the gas is heated during reionization, it hydrodynamically responds to the resulting increase in its temperature and pressure by expanding \citep{Hui97,Puchwein2023}. The more heat injected, the more the gas expands, erasing more small-scale structure. In our models we parametrised this effect with the cumulative heat injected per proton by a given redshift ($u_0$) \citep{Nasir16}. The exact redshift range of $u_0$ parameters is the same as in \citep{Boera2019}.

The small-scale structure in the gas could further be affected by the free-streaming of non-standard dark matter models such as WDM. The lighter the mass of a thermal relic WDM particle, the longer the particles will free-stream, from when they decouple from the thermal bath until they become non-relativistic. The longer this time the larger the scales affected, and the stronger the suppression in the small-scale power. This is shown in the right-hand side panel of Fig.~\ref{fig:p1d_models_2}, where the proxy for the free-streaming scale used is the inverse of the particle mass, $m_{\rm WDM}^{-1}$.

\subsection{Mass resolution and boxsize}

Since our models are built from the results of hydro-dynamical simulations it is important to understand whether the results of these simulations are numerically converged. Two main factors limit this convergence \citep{bolton17,lukic15,Doughty2023} -- the size of the simulated box limits the number of large-scale modes and affects the convergence on large scales; and the mass or particle resolution of the simulation limits the smallest resolved scale.

\begin{figure*}
    \includegraphics[scale=0.55]{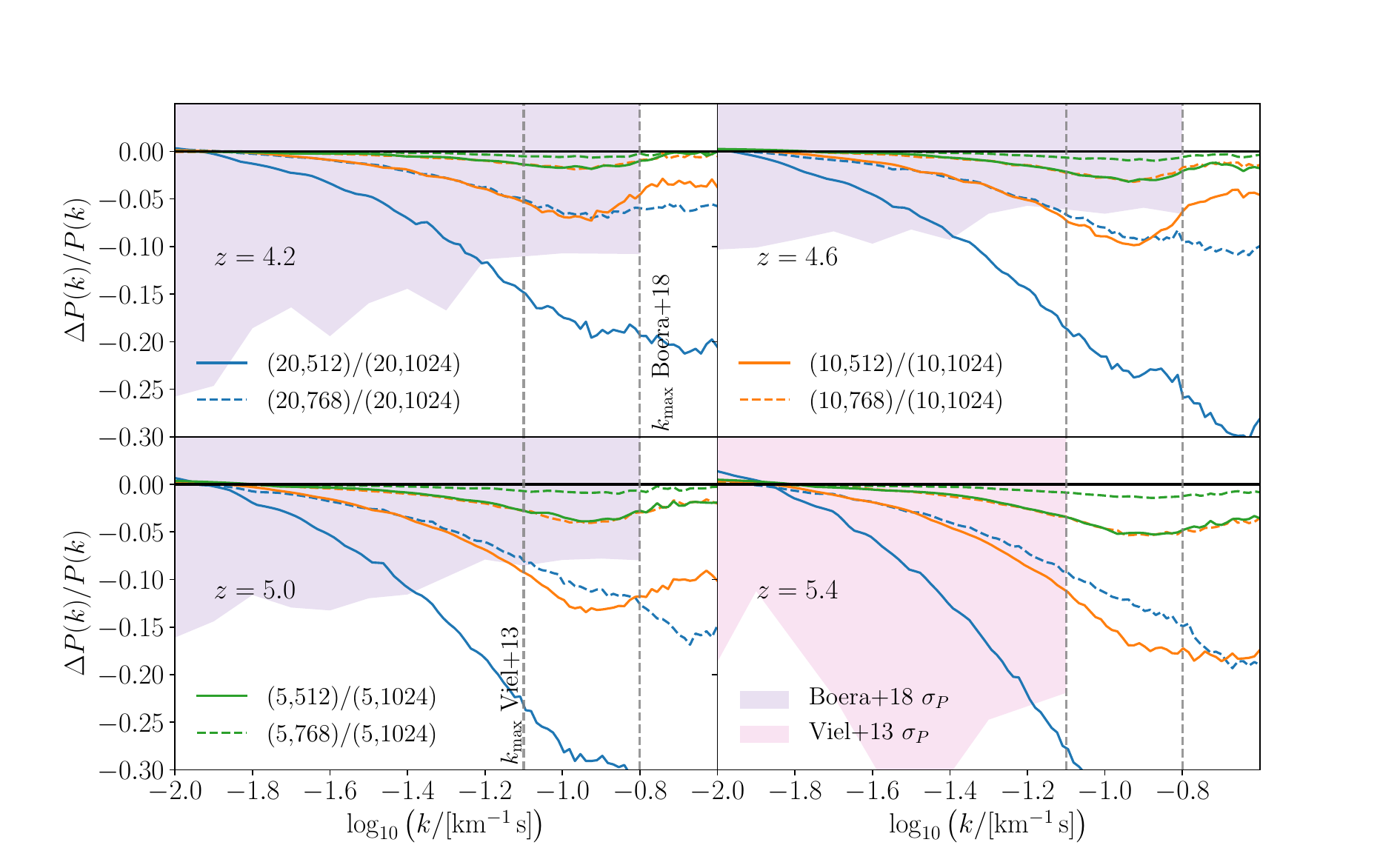}
    \caption{The effect of mass resolution in the simulations, shown as a flux power spectrum decrement as a function of wavenumber for simulations of varying particle numbers. The mass resolution decrement of the flux power spectrum is largest for the lowest resolution simulations (blue-solid) and smallest for the highest resolution simulations (green-dashed). The decrement as a function of mass resolution decreases, indicating convergence. The fiducial grid of simulations (20,1024) used in this work is converged at the 5-10\% level at $k=0.2\;\skm$. The default mass resolution correction uses models with higher mass resolution (10,1024) that are converged at 2-5\% at $k=0.2\;\skm$. The shaded regions show the observational $1\sigma$ uncertainty on the flux power spectrum from \citep{Viel13wdm} (pink) and \citep{Boera2019} (violet). The vertical dashed lines indicate the $k_{\rm max}$ of different data sets.
    }    
    \label{fig:mass_resolution}
\end{figure*}

We have supplemented our simulation suite with additional calibration runs varying the size of the simulated box at fixed mass resolution. Our fiducial grid of simulations uses a box size of 20 $\Mpch$. We have applied the splicing correction, \citep{McDonald05}, using the 40 $\Mpch$ box with the same resolution as L20-ref, which results in a correction of the level of $\leq 3$\% on the 1D flux power spectrum in the low-$k$ regime. We have further verified that at the scales of interest for the analysis of \citep{Boera2019} data, further corrections using 80 and 160 $\Mpch$ box sizes were negligible. This was not an unexpected result, and has been observed in several previous studies \citep{Viel13wdm,Irsic2017b,Boera2019}.

Of more importance for the studies of the small-scale 1D flux power spectrum, is the mass resolution of the simulations ($R_s$). The grid of simulations was run with the fiducial gas mass resolution of $9.97\times10^{4}$ $h^{-1}\,M_{\odot}$, corresponding to $2$x$1024^3$ baryon and dark matter particles. These models are converged at the 5-10\% at the smallest scales used in the analysis. We have complemented these models with additional simulations varying the number of simulated particles at different fixed box sizes. 

Fig.~\ref{fig:mass_resolution} shows the 1D flux power spectrum decrements between different models. The poorer the mass resolution of the simulation the larger the suppression of the small-scale flux power spectrum relative to a higher resolution simulation. The mass resolution correction is larger at higher redshifts, and at smaller scales, in agreement with previous results in the literature (e.g. \citep{bolton17,Doughty2023}). The mass resolution correction ($R_s$) and the 1D flux power decrements shown in in Fig.\ref{fig:mass_resolution} are connected as $R_s^{-1} = 1+\Delta P/P$. The grid of our simulations at the resolution of (20,1024) was corrected for the residual mass resolution with (10,1024) model (R-set; see Table~\ref{table:summary_sims}), corresponding to gas mass resolution of $1.25\;\times 10^{4}$ $h^{-1}\,M_{\odot}$. Additional correction due to higher resolution simulations (e.g. (5,1024)) adds less than a few percent to the total mass resolution correction.

\section{Results}
\label{sec:results}

\begin{figure*}
    \centering
    \includegraphics[scale=0.45]{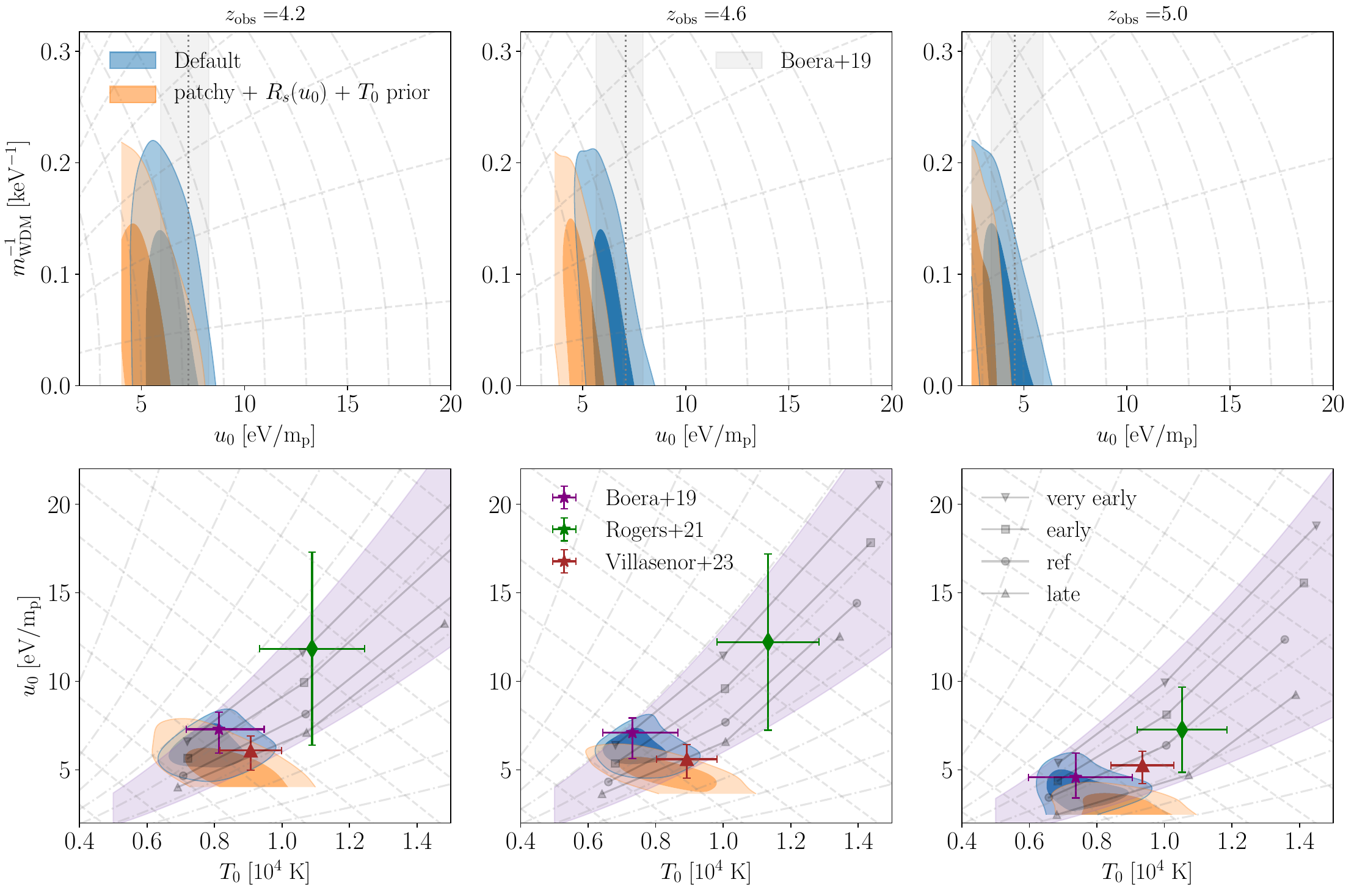}
    \caption{The 2D posterior distributions of the best-fit analysis for the 1D flux power spectrum measurements of \citep{Boera2019} using UVES/HIRES quasar spectra. The blue contours show the default analysis, and the orange contours show the analysis that captures our best knowledge of the thermal history (Sec.~\ref{sec:Thermal_history}), inhomogenous reionization (Sec.~\ref{sec:patchy_reionization}) and mass resolution corrections (Sec.~\ref{sec:thermal_mass_resolution}). The three columns correspond to the three different redshifts of $z=4.2,4.6$ and $5.0$ (from left to right). The bottom row shows the contours in the thermal parameter space, with the violet band shows the envelope around the physically motivated simulations, shown as gray points (squares, circles and triangles). This band serves as a prior in the thermal parameter space in the default model. The coloured points correspond to the measurements in the literature from the same data set: from \citep{Boera2019} (purple); \citep{Rogers2021} (in green); and \citep{Villasenor2022b} (in red). The top panels show the 1 and 2 $\sigma$ contours in the parameter space of free-streaming and pressure smoothing (heat injection). The vertical dotted line and surrounding gray band indicate the best-fit measurements of \citep{Boera2019}. The intersecting gray dashed and dot-dashed lines show typical degeneracy axes between the parameters. The cutoff at small $u_0$ and small $T_0$ values comes from the implicit prior imposed by the extent of the grid of models (see text for details).
    }    
    \label{fig:panel_2d_wdm_thermal}
\end{figure*}

The new results on the free-streaming of warm dark matter are summarized in Fig.~\ref{fig:panel_2d_wdm_thermal}. The six panels show the 2D posteriors for three redshift bins of the data \citep{Boera2019}, with the redshift label referring to the label of the thermal parameters that are independent in each redshift bin. The bottom row shows the constraints in the thermal parameter space of gas temperature and pressure smoothing (through the proxy of cumulative injected heat), whereas the top row shows the constraints spanning the parameter space of pressure smoothing and free-streaming.

The fiducial analysis choice assumes priors on the thermal history in the $u_0-T_0$ plane as an envelope around our fiducial grid of simulations (see below). We also assume Planck \citep{Planck2020} priors on CDM cosmology parameters $(\sigma_8,n_s)$. For the default analysis we use mass resolution correction using a fiducial thermal history with CDM cosmology (R-set; see Table~\ref{table:summary_sims}). We also do not include any correction due to inhomogenous reionization. These assumptions were chosen as our reference analysis in order to facilitate better comparison with previous analysis. The additional work presented in this paper which includes patchy correction, thermal dependence of the mass resolution correction ($R_s(u_0)$) and observationaly informed thermal priors ($T_0$ prior) is also shown in Fig.~\ref{fig:panel_2d_wdm_thermal} (orange contours) and discussed in more detail in subsections below.

Our measurements of the thermal state of the gas largely agree with independent measurements in the literature \citep{Boera2019,Rogers2021,Villasenor2022b} within 1-2$\sigma$. The data prefers a slightly colder temperature at mean density of $T_0 = 8,000$ (7,500; 7,800) $\mathrm{K}$ at redshift $z=4.2$ (4.6; 5.0) as a best-fit (see Table~\ref{table:summary_mcmc}). At the same time the cumulative heat injected is constrainted to be $u_0 = 7.2$ (6.8; 5.2)$\;\mathrm{eV/m_p}$ between redshifts $4.2$ and $12.0$ (4.6 and 12.0; 6.0 and 13.0). The result is consistent with the analysis of \citep{Boera2019}. However, models with slightly hotter temperature consistent with \citep{Gaikwad20,Gaikwad21} and less pressure smoothing \citep{kulkarni19} are within the $2\sigma$ contours. 

The measurements of effective optical depth, $\tau_{\rm eff}$, from the flux power spectrum are also consistent with direct observations of the transmitted flux \citep{Becker13,Bosman2021}. The derived measurement of the mean transmitted flux at $z=5.0$ is $\langle F_{\rm Ly\alpha} \rangle = 0.1764_{-0.0171}^{+0.0177}$. This is consistent at $1-2\sigma$ with the measurement of \citep{Bosman2021} of $\langle F_{\rm Ly\alpha} \rangle = 0.1581_{-0.0089}^{+0.0082}$ which used almost four times the number of sightlines compared to \citep{Boera2019}. 

This analysis also varies the power-law of the temperature-density relation ($\gamma$) as a free parameter in each redshift bin. The data, however, are not constraining this parameter well and its posterior is dominated by the prior. This result was also found in previous studies of high redshift \lya{} forest data (e.g. \citep{Irsic2017b,Boera2019}).

The panels at the bottom of Fig.~\ref{fig:panel_2d_wdm_thermal} also show the $u_0-T_0$ combinations of hydro-dynamical simulations as gray markers (L20; see Table~\ref{table:summary_sims}). The thermal and reionization histories were chosen to bracket the observed flux distribution of high redshift quasar spectra \citep{bosman18}, as well as the electron optical depth inferred from the Cosmic Microwave Background (CMB) as reported by Planck \citep{Planck2020,Belsunce2021}. Through the post-processing technique described in Sec.~\ref{sec:simulations} the likelihood is able to sample the full span of the $u_0-T_0$ parameter space on a (non-uniform) grid, however in order to avoid unphysical parts of the $u_0-T_0$ parameter space we consider a prior defined as an envelope around the simulations' results (indicated in Fig.~\ref{fig:panel_2d_wdm_thermal} by the gray band).

\subsection{Degeneracy axes}

The simulated models exhibit a tight correlation between the IGM temperature at a given time, and the integrated injected heat up until that time. The positive correlation between the thermal parameters (dot-dashed lines in bottom panel of Fig.~\ref{fig:panel_2d_wdm_thermal}) can be well described by $u_0 \propto T_0^{1.7}$, and the parameter anti-correlation (dashed lines in bottom panel of Fig.~\ref{fig:panel_2d_wdm_thermal}) is well described by $u_0 \propto -T_0$. 
The anti-correlation also indicates the degeneracy axis we would expect from the measurement of the 1D flux power spectrum -- at a given observed redshift the flux power suppression can be explained by either higher injected heat, and therefore a larger pressure smoothing scale; or it can be explained by a higher temperature and therefore larger thermal broadening. The \lya{} forest provides constraints in the direction perpendicular to that degeneracy axis, along the direction of the positive correlation between $u_0$ and $T_0$. 

Similarly to the degeneracy between the thermal broadening and pressure scales, we observe a correlation between the pressure smoothing and the free-streaming scales, as shown in the top panels of Fig.~\ref{fig:panel_2d_wdm_thermal}. The vertical black dashed line and gray shaded region, indicate measurements of the cumulative injected heat, $u_0$, in a CDM analysis of \citep{Boera2019}. A negative correlation (dot-dashed lines in top panel of Fig.~\ref{fig:panel_2d_wdm_thermal}) between the two smoothing scales can be understood as a consequence of both physical mechanisms reducing the small-scale power of the 3D density field. The pressure smoothing scale is typically described as an exponential suppression of the power, $P_{\rm g} \sim P_{\rm m} \exp\left(-k^2 \lambda_F^2\right)$ \citep{Puchwein2023,Hui98}, at a typical filtering scale $\lambda_F$. The larger the heat injected into the gas, the more the gas expands due to the pressure, resulting in a positive correlation between the filtering scale and the injected heat $u_0$. Such a relation was explored in the simulations of \citep{Boera2019}, where it was found that $\lambda_F \sim 20\;\mathrm{ckpc} \times \sqrt{1 + 2 u_0/(1 eV/m_p)}$. 

Equivalently, the warm dark matter transfer function can be approximated by $T_{\rm WDM} \sim \left[1 + (\alpha k)^{2\mu}\right]^{-5/\mu}$, with $\mu = 1.12$ and the typical free-streaming scale, $\alpha = 70\;\mathrm{ckpc} \times \left(m_{\rm WDM}/(1 \mathrm{keV})\right)^{-1.11}$, given by \citep{Viel13wdm}.

The total power suppression in the 3D field on small scales, is a product of both the pressure smoothing and free-streaming transfer functions. Expanding the product in powers of $k$, the lowest scale dependent coefficient scales as $\propto k^2$, with the amplitude of $c_2^2 = \lambda_F^2 + 10 \alpha^2$, where we have approximated $\mu \sim 1$. The anti-correlation between pressure smoothing and the free-streaming that we observe in the data are driven by being sensitive to the total shape of the suppression, thus $c_2^2 = \mathrm{constant}$. This can be interpreted as the smoothing being driven by either higher pressure smoothing or larger free-streaming length, and is shown as dot-dashed gray lines in Fig.~\ref{fig:panel_2d_wdm_thermal}.

Whereas the shape of the power spectrum suppression is poorly constrained by the current data, the data is able to constrain the scale where the suppression occurs -- shown as dashed lines in Fig.~\ref{fig:panel_2d_wdm_thermal}. We estimate this positive correlation between the parameters (dashed lines in top panel of Fig.~\ref{fig:panel_2d_wdm_thermal}) by matching the scale where the pressure smoothing and free-streaming transfer functions equal one half (e.g. $T_{\rm WDM}(k_{\rm 1/2} = 1/2$, or $P_{\rm WDM}(k_{\rm 1/2})=1/4$). The two scales are given by $k_{\rm 1/2}^{g} = \sqrt{2\log{2}}/\lambda_F$, and $k_{\rm 1/2}^{\rm WDM} = (-1 + 2^{\mu/5})^{\mu/2}/\alpha$. Equating the two leads to a relation $m_{\rm WDM}^{-2.22} \propto 1 + 2 u_0/(1 \mathrm{eV}/m_p)$, that defines the directional axis along which the \lya{} forest data gives the tightest constraints.

\begin{figure*}
    \centering
    \includegraphics[width=1.0\textwidth]{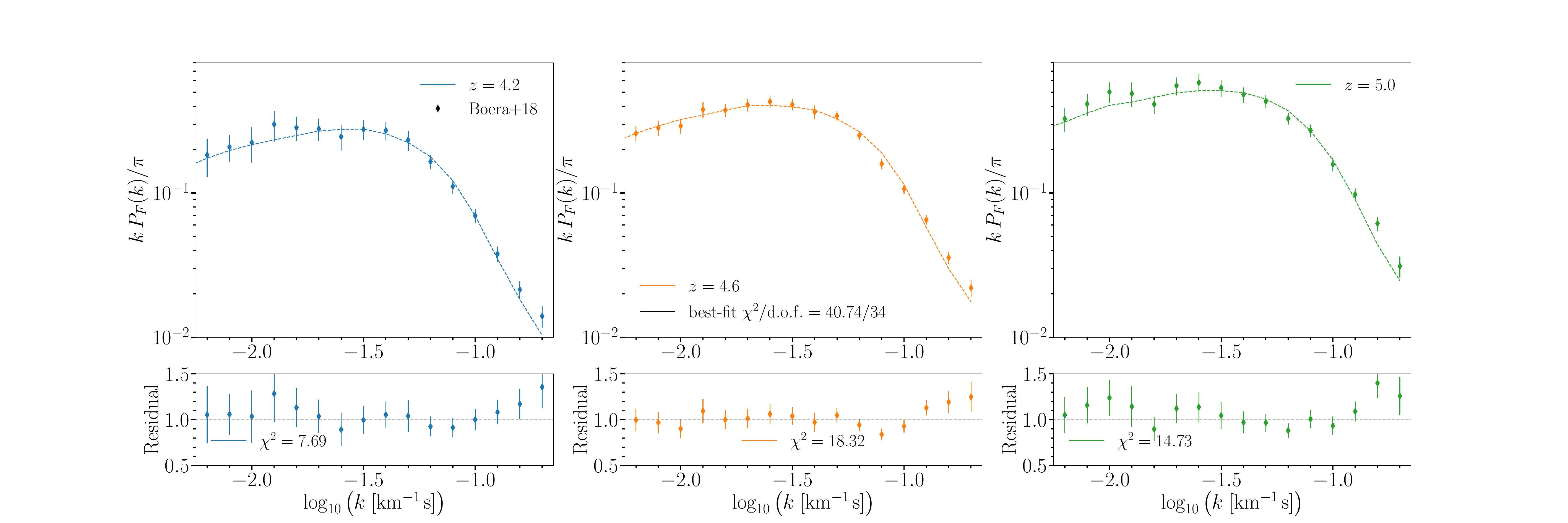}
    \caption{The best-fit model compared to the data \citep{Boera2019}. The three panels correspond to three redshift bins, with the bottom panels showing the residuals of the data over the model. The total $\chi^2$ is $40.7$ with 34 degrees of freedom. The data were compared to a simulation based model that varies three thermal parameters and mean transmission independently in each redshift bin ($\tau_{\rm eff},T_0,\gamma,u_0$) and three cosmology parameters ($\sigma_8$,$n_s$) and ($m_{\rm WDM}$) (see text for details).
    }    \label{fig:bestfit_wdm_main}
\end{figure*}

\subsection{Best-fit model}

Fig.~\ref{fig:bestfit_wdm_main} shows 1D flux power spectrum corresponding to the best-fit model over-plotted on the data. The model fits the data reasonably well, with a total $\chi^2$ of 40.7 and 34
degrees of freedom (see Table~\ref{table:summary_mcmc}). Furthermore, the model is in excellent agreement with the data up to $k\sim 0.1\;\skm$, and describes the position and shape of the flux power spectrum suppression on small scales. To illustrate this we can compare the model that is fit to all the data points and re-evaluate the $\chi^2$ for the points up to $k<0.1\;\skm$. In this case the fit gives $\chi^2$ of 20.4 with 20
degrees of freedom. All three redshift bins show an increase in the measured power relative to the model at $k>0.1\;\skm$. This indicates a possible shortcoming of the model on the smallest scales, or else a signal in the data that is not part of the model.

The best-fit model excludes $m_{\rm WDM} < 5.7\;\mathrm{keV}$ (95\% C.L.) and provides the tightest constraints on the thermal relics WDM particle mass to date (see Table~\ref{table:summary_mcmc}). The model constraints exclude masses of $3.73\;\mathrm{keV}$ and $3.18\;\mathrm{keV}$ at $3\sigma$ and $5\sigma$, effectively excluding the much discussed $3\;\mathrm{keV}$ WDM model (e.g. \citep{Adhikari2017,Enzi2021}) at more than a $5\sigma$ confidence level.

\begin{figure}
    \centering
    \includegraphics[width=0.48\textwidth]{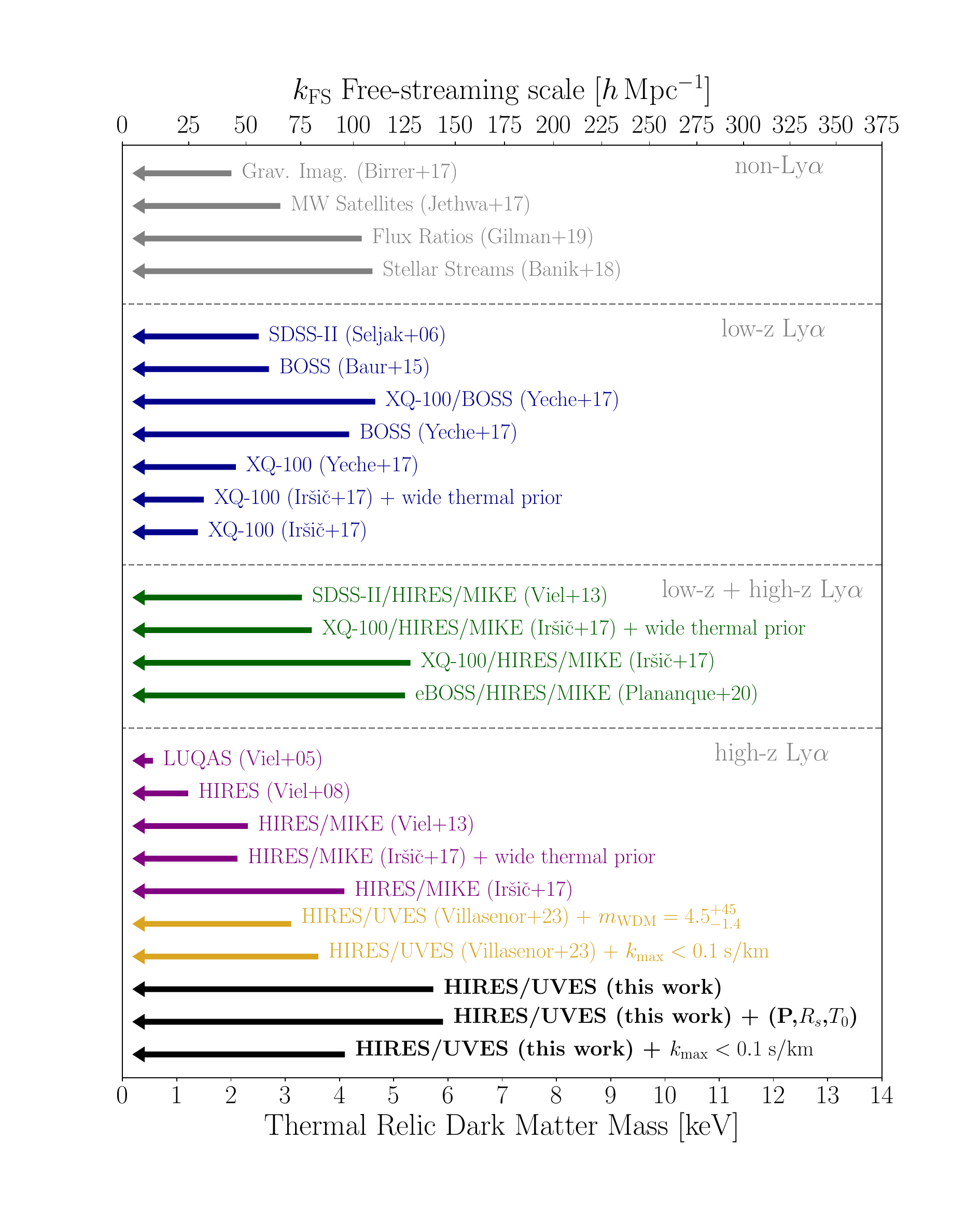}
    \caption{The $2\sigma$ constraints on the thermal relic warm dark matter mass. The arrows indicate the exclusion limits on the WDM particle mass in $\mathrm{keV}$. The bottom panel shows a compilation of constraints from high redshift \lya{} forest 1D flux power spectrum. The black arrows at the very bottom indicate the results of this study, for three different analysis choices pertaining the measured flux power at highest wavenumbers (default, $(P,R_s,T_0)$ that represents corrections due to patchy reionization, thermal dependence on the mass resolution and independent $T_0$ prior, and $k_{\rm max}<0.1\;\skm$ data scale cut analysis). The resulting lower bounds on the WDM particle mass are stronger or comparable to those previously published in the literature, including studies that combined low- and high- $z$ \lya{} forest data to increase the redshift lever arm (middle panels). The top panel shows a compilation of results from non-\lya{} studies.
    }    \label{fig:wdm_constraints}
\end{figure}

\subsection{Improvement on WDM constraints}

In Fig.~\ref{fig:wdm_constraints} we compare the results of this work to the existing constraints on the WDM mass from the literature. The main result of this work results in a WDM mass bound that excludes WDM masses below $m_{\rm WDM} < 5.7\;\mathrm{keV}$ at $2\sigma$ confidence level. It provides improved constraints on WDM mass coming from the matter power spectrum suppression in the \lya{} forest analyses \citep{Irsic2017b,palanque20} as well as non-\lya{} constraints such as the flux ratios of strong lensed systems \citep{Gilman2020} and stellar streams in the Milky-Way \citep{Banik2018}.

The new constraint is stronger than the studies using low-$z$ \citep{Yeche2017} or a combination of low-$z$ and high-$z$ \citep{Viel13wdm,Palanque13} \lya{} data, especially when comparing to similar choices in the thermal history priors. The new data is in fact producing a strong enough constraint that, even when relaxing the prior on the astrophysical parameters, the WDM mass bound remains stronger or competitive with past studies that used strong priors on the e.g. temperature evolution with redshift \citep{Irsic2017b,palanque20}.

In the regime of the high redshift \lya{} forest analysis, the current analysis tightens the constraint on the WDM particle mass compared to previous analyses. In comparison to older analyses using HIRES/MIKE data \citep{Viel13wdm,Irsic2017b} we see an improvement in the number of the observed quasar spectra by almost a factor of 2 \citep{Boera2019}. For a factor of 2 improvement in the number of sightlines, we would expect the uncertainty on the flux power spectrum to improve by $\sim1/\sqrt{2}$, at least in the limit that statistical uncertainty dominates the error budget. From Fig.~\ref{fig:mass_resolution} we see that this is indeed the case in the high-$k$ regime of the data that is most sensitive to the free-streaming effect of WDM. In fact, in linear theory the sensitivity to the the WDM mass scales as $P_{\rm L,wdm}/P_{\rm L,CDM}\sim m_{\rm wdm}^{20} k^{-20}$ in the limit of $k\gg 14\;(m_{\rm wdm}/1\;\mathrm{keV})\;\mathrm{Mpc^{-1}}$.

However, the non-linear mapping between the linear density field and the non-linear flux field is complex. For a range of redshifts ($4.2<z<5.0$) and scales ($0.01<k/\mathrm{[km^{-1}\,s]}<0.2$) considered, the flux power spectrum suppression in our simulations (L20-ref) approximately scales as 
\begin{equation}
    \frac{P_{\rm F,wdm}}{P_{\rm F,cdm}} \sim 
    \begin{cases}
    1 - 0.1\left(\frac{1+z}{5}\right)^4 \left(\frac{k}{0.1}\right)^{\frac{3}{4}} \left(\frac{m_{\rm wdm}}{4}\right)^{-1}, & m_{\rm wdm} > 3\;\mathrm{keV}\\
    1 - 0.1\left(\frac{1+z}{5}\right)^3 \left(\frac{k}{0.1}\right)^{\frac{1}{2}} \left(\frac{m_{\rm wdm}}{4}\right)^{-\frac{3}{2}}, & m_{\rm wdm} < 3\;\mathrm{keV},
    \end{cases}
\end{equation}
with line-of-sight wavenumber $k$ in units of $[\mathrm{km^{-1}\,s}]$ and $m_{\rm wdm}$ in units of $[\mathrm{keV}]$. For higher WDM masses, the flux power suppression due to WDM increases rapidly with redshift, but only linearly with the WDM particle mass. The scaling changes at around the WDM mass of $3\;\mathrm{keV}$, when the scaling with mass becomes stronger, and the redshift dependence slightly weaker. The wavenumber dependence is roughly the same, and not dominant in this range of scales. The scaling is only approximately valid at a fixed thermal history (L20-ref), and the transition between the two scales, as well as the power-law dependencies, can vary across thermal histories. However, at the higher WDM mass limit the sensitivity to the particle mass increases with redshift relatively quickly in the redshift range of the data, improving the linear sensitivity to the mass. As a result the constraining power on WDM mass improves by more than a factor of $\sim 1/\sqrt{2}$.

\begin{table*}
\centering
\caption{List of different models used in the analysis with their corresponding best-fit warm dark matter constraints. The table shows the name of the model and the resulting $2\sigma$ lower bound on the WDM particle mass ($m_{\rm WDM}$), along with best-fit values of the thermal parameters at $z=4.6$ for the effective optical depth ($\tau_{\rm eff}$), gas temperature at mean density ($T_0$), the slope of the temperature-density relation ($\gamma$) and the cumulative injected heat ($u_0$). For the model where extra instrumental noise in the data was modelled with a free parameter, the best-fit value is shown as well ($A_{\rm noise}$). The last column displays the best-fit $\chi^2$ value and the degrees of freedom.
}
\label{table:summary_mcmc}
\begin{tabular}{lccccccc}
\hline
Name    & $m_{\rm WDM}\;\mathrm{[keV]}\;(2\sigma)$ & $\tau_{\rm eff}(z=4.6)$ & $T_0(z=4.6)\;\mathrm{[10^4\;K]}$ & $\gamma(z=4.6)$ & $u_0(z=4.6)\;\mathrm{[eV/m_p]}$ & $A_{\rm noise}(z=4.6)$ & $\chi^2/{\rm dof}$ \\
\hline 
Default & $>5.72$ & $1.502_{-0.061}^{+0.061}$ & $0.743^{+0.041}_{-0.075}$ & $1.35^{+0.24}_{-0.19}$ & $6.19_{-0.68}^{+0.68}$ & - & $40.7/34$ \\ 
\hline
$k_{\rm max} < 0.1\;\skm$ & $>4.10$ & $1.501^{+0.060}_{-0.074}$ & $0.840_{-0.340}^{+0.095}$ & $1.28_{-0.28}^{+0.09}$ & $8.91_{-5.26}^{+1.57}$ & - & $10.2/20$  \\ 
$A_{\rm noise}$ & $>3.91$ & $1.458^{+0.053}_{-0.074}$ & $0.966_{-0.466}^{+0.156}$ & $1.23_{-0.23}^{+0.06}$ & $5.93_{-2.28}^{+0.38}$ & $1.12^{+0.49}_{-0.29}$ & $18.4/31$  \\ 
$T_0$ prior & $>5.85$ & $1.494^{+0.062}_{-0.077}$ & $0.770^{+0.110}_{-0.120}$ & $1.31_{-0.31}^{+0.10}$ & $6.50^{+1.00}_{-1.60}$ & - & $47.6/34$  \\ 
$R_s(u_0)$ mass resolution & $>4.44$ & $1.531^{+0.073}_{-0.064}$ & $0.617_{-0.118}^{+0.007}$ & $1.38^{+0.28}_{-0.13}$ & $7.90^{+1.70}_{-2.30}$ & - & $30.7/34$  \\ 
patchy reion.& $>5.10$ & $1.486^{+0.058}_{-0.068}$ & $0.686^{+0.046}_{-0.080}$ & $1.33^{+0.17}_{-0.26}$ & $5.32^{+0.58}_{-0.52}$ & - & $41.0/34$ \\ 
\hline
$R_s(u_0)$ + $T_0$ prior & $>4.24$ & $1.473^{+0.056}_{-0.076}$ & $0.83^{+0.11}_{-0.11}$ & $1.28^{+0.09}_{-0.28}$ & $5.53^{+0.73}_{-1.2}$ & - & $39.4/34$ \\
patchy + $R_s(u_0)$ + $T_0$ prior & $>5.90$ & $1.450^{+0.051}_{-0.070}$ & $0.828^{+0.098}_{-0.098}$ & $1.26^{+0.08}_{-0.26}$ & $4.87^{+0.52}_{-0.71}$ & - &  $40.8/34$\\
\hline
\end{tabular}
\end{table*}

\subsection{Thermal history}
\label{sec:Thermal_history}

The signal at high-$k$ in the 1D \lya{} forest flux power spectrum depends on the thermal parameters. The fiducial priors on the thermal history limit the possible combinations in the $u_0-T_0$ plane to the volume of physically motivated simulation results \citep{Puchwein2023}.

\begin{figure}
    \includegraphics[width=0.5\textwidth]{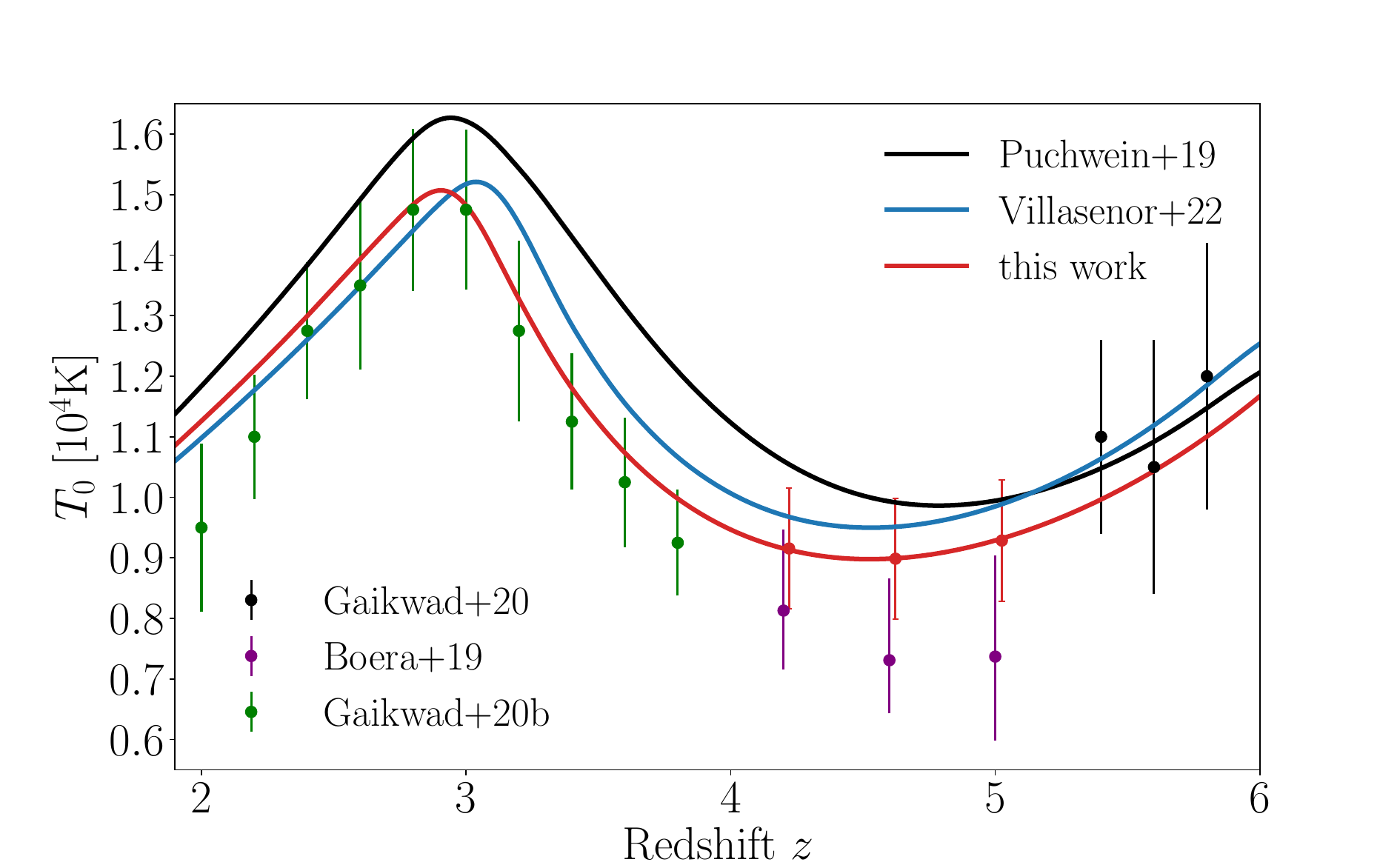}
    \caption{The thermal evolution of $T_0(z)$ for various models, compared to the independent measurements of \citep{Gaikwad20} (high-$z$), \citep{Gaikwad21} (low-$z$) and temperature measurements of \citep{Boera2019}. The models shown are that of \citep{Puchwein19} (in black; our reference simulation run), \citep{Villasenor2022a} (in blue) and a new fit to \citep{Gaikwad20,Gaikwad21} data (in red). The new fit was obtained by rescaling the model of \citep{Puchwein19}, and serves as a prior in the analysis shown in Fig.~\ref{fig:effect_thermal_priors}, with prior values shown as red circles and error bars at $z=4.2,4.6,5.0$ with values of 9155.5, 8986.5 and 9286.5 K respectively. The uncertainty propagated in the prior is 1,000 K at each of the redshifts.
    }    
    \label{fig:T0z}
\end{figure}

A different approach would be to instead apply a prior based on independent measurements of the thermal history, for example using the measurements of $T_0(z)$ from different data-sets and different statistical methods. To achieve that we use improved and precise measurements of $T_0(z)$ that span the redshift range $z<3.8$ \citep{Gaikwad20} and $z>5.2$ \citep{Gaikwad21}. In order to predict viable models in the redshift range covered by our data ($4.2<z<5.0$) we rescale and shift the photo-heating and photo-ionization rates of our fiducial thermal history model \citep{Puchwein19}. A similar methodology was employed in \citep{Villasenor2022a} in order to fit the flux power spectra measurements over a range of redshifts. Fig.~\ref{fig:T0z} shows the two models from the literature, as well as our new fit calibrated directly against $T_0(z)$ measurements. The best-fit prefers slightly higher temperatures in the redshift range $4.2<z<5.0$ than the measurements of \citep{Boera2019} using the flux power spectra data used in this work. In order to construct informative priors on $T_0(z)$ parameters in our model, we use the best-fit values of the new thermal history as central points of a Gaussian distribution at each redshift, with a fixed standard deviation of 1,000 K. While the standard deviation is somewhat arbitrarily chosen, it roughly matches the typical uncertainty found in more recent works \citep{Gaikwad20,Gaikwad21}. Even if a realistic uncertainty is slightly lower at lower redshift, and slightly higher at higher redshift, due to the decreasing numbers of quasar spectra available, this should not impact the main conclusion of this exercise, which is to highlight the effect of independent, observationally informed thermal priors.

\begin{figure*}
\centering
    \includegraphics[width=0.45\textwidth]{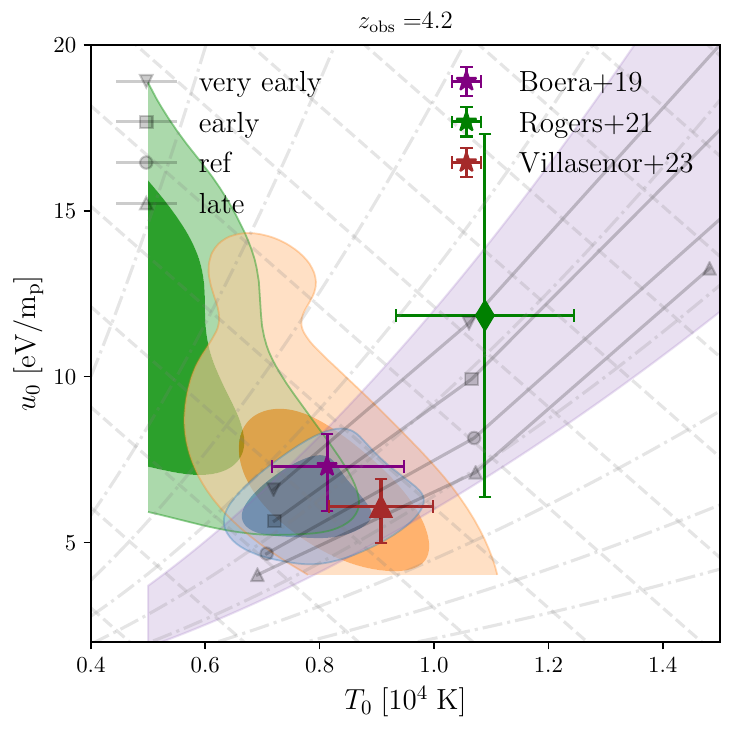}
    \includegraphics[width=0.45\textwidth]{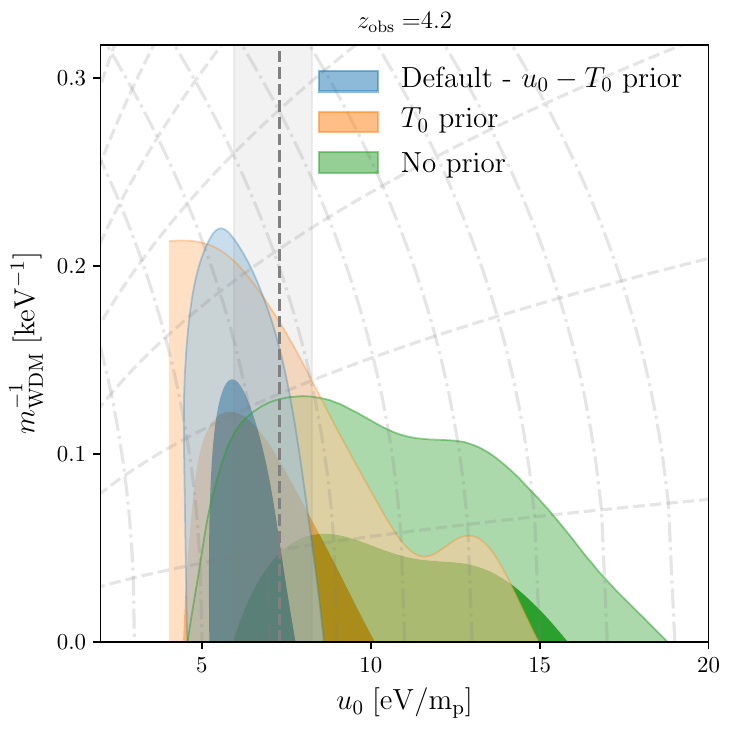}
\caption{Effect of thermal priors on the posterior. The two panels show 2D posterior distributions for redshift $z=4.2$ in the plane of temperature and heat injection (left) and warm dark matter mass and heat injection (right). In the thermal parameter space (left), the default analysis (blue contours) uses thermal priors that envelop the simulations (the envelope is shown as a violet shaded area). A similar result can be obtained by instead imposing a $T_0(z)$ prior (orange contours) using independent temperature measurements \citep{Gaikwad20,Gaikwad21}. The warm dark matter particle mass constraints get slightly stronger if a temperature prior is used instead of the envelope prior in the $u_0-T_0$ plane. As a reference we also show an analysis without imposing any thermal priors (green contours). The vertical shaded band on the left panel indicates measurement of $u_0$ from \citep{Boera2019}.
}
\label{fig:effect_thermal_priors}
\end{figure*}

The results of the analyses using different thermal prior choices are shown in Fig.~\ref{fig:effect_thermal_priors}. The fiducial model (green contours) uses thermal priors in the form of an envelope around the simulations (gray band). Replacing these priors by simpler priors on $T_0$ at each redshift results in slightly more elongated constraints on the $u_0-T_0$ plane (orange contours), with the posterior expanding along the degeneracy direction. The mean and best-fit of the posterior however, change only marginally compared to the standard analysis. Even though the posterior in the $u_0$ direction expands slightly, the posterior on $m_{\rm WDM}^{-1}$ remains roughly the same at the $2\sigma$ level, resulting in a very similar constraint on the WDM mass of $m_{\rm WDM} > 5.85\;\mathrm{keV}$ $(2\sigma)$ compared to the default analysis choice of thermal priors. The main difference is that the thermal priors have now been informed by the measured $T_0$ evolution with redshift from other observational studies, rather than by our suite of hydro-dynamical simulations. The model with the $T_0$ prior (see Table~\ref{table:summary_mcmc}) excludes low WDM masses of $3.75\;\mathrm{keV}$ and $3.21\;\mathrm{keV}$ at $3\sigma$ and $5\sigma$, respectively.

Fig.~\ref{fig:effect_thermal_priors} also illustrates the effect of not imposing any thermal priors on the analysis. This resulting posterior (blue contours) is shifted to lower IGM temperatures, and relatively higher values of the cumulative injected heat. This part of the thermal parameter space is unphysical as we expect $u_0$ and $T_0$ to be correlated for physically reasonable IGM heating scenarios. Rather counter-intuitively the constraints on the WDM mass become much stronger if we impose no thermal prior. This result can be understood by considering degeneracy axis along which the posterior distributions move. Colder temperatures and enhanced amount of pressure smoothing leave much less room for a WDM model to accommodate the amount of flux power spectrum suppression in the data. Therefore, models with strong WDM suppression are excluded more strongly.

\subsection{Effect of small-scale peculiar velocity}
\label{sec:pecvel_effects}

\begin{figure*}
\centering
    \includegraphics[width=0.45\textwidth]{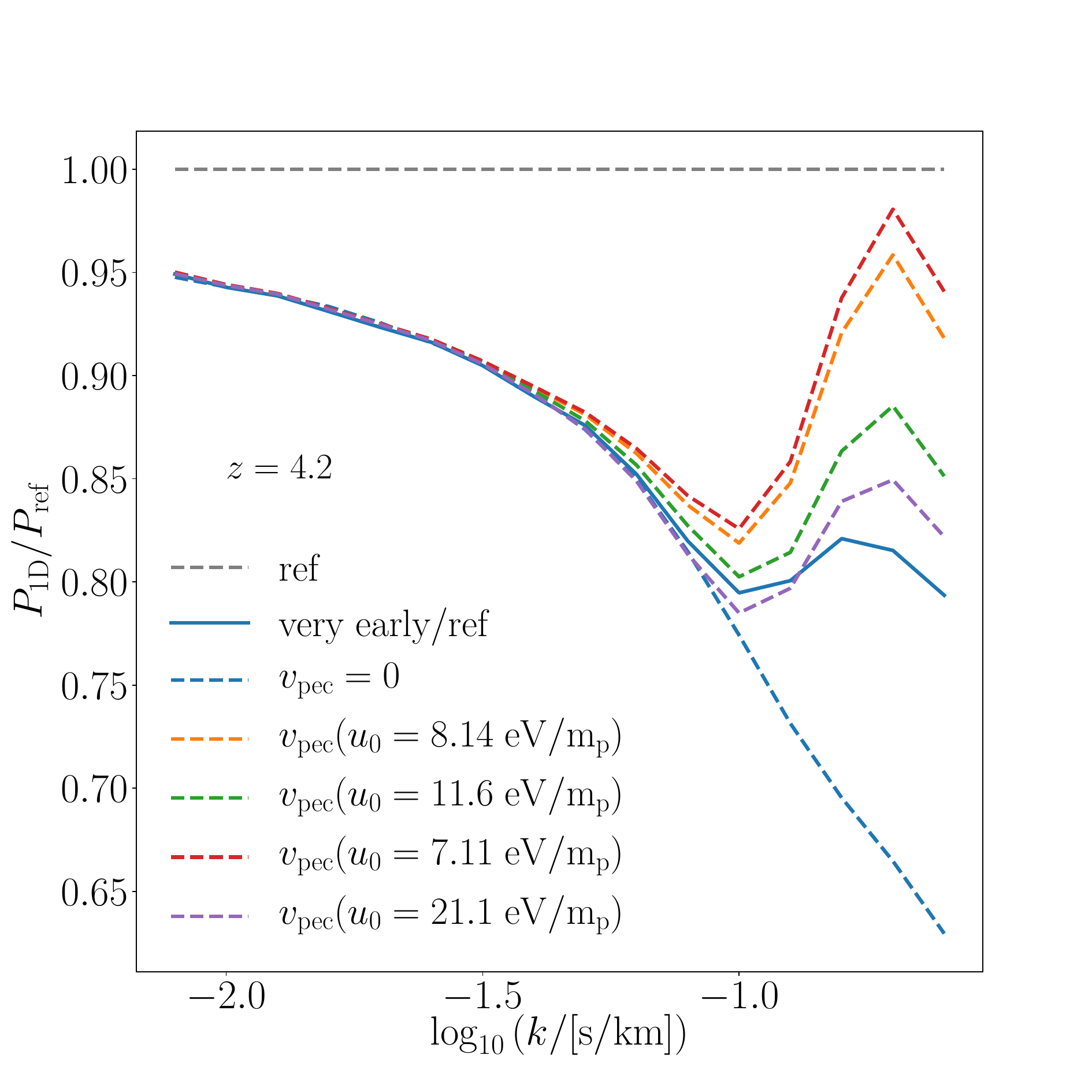}
    \includegraphics[width=0.45\textwidth]{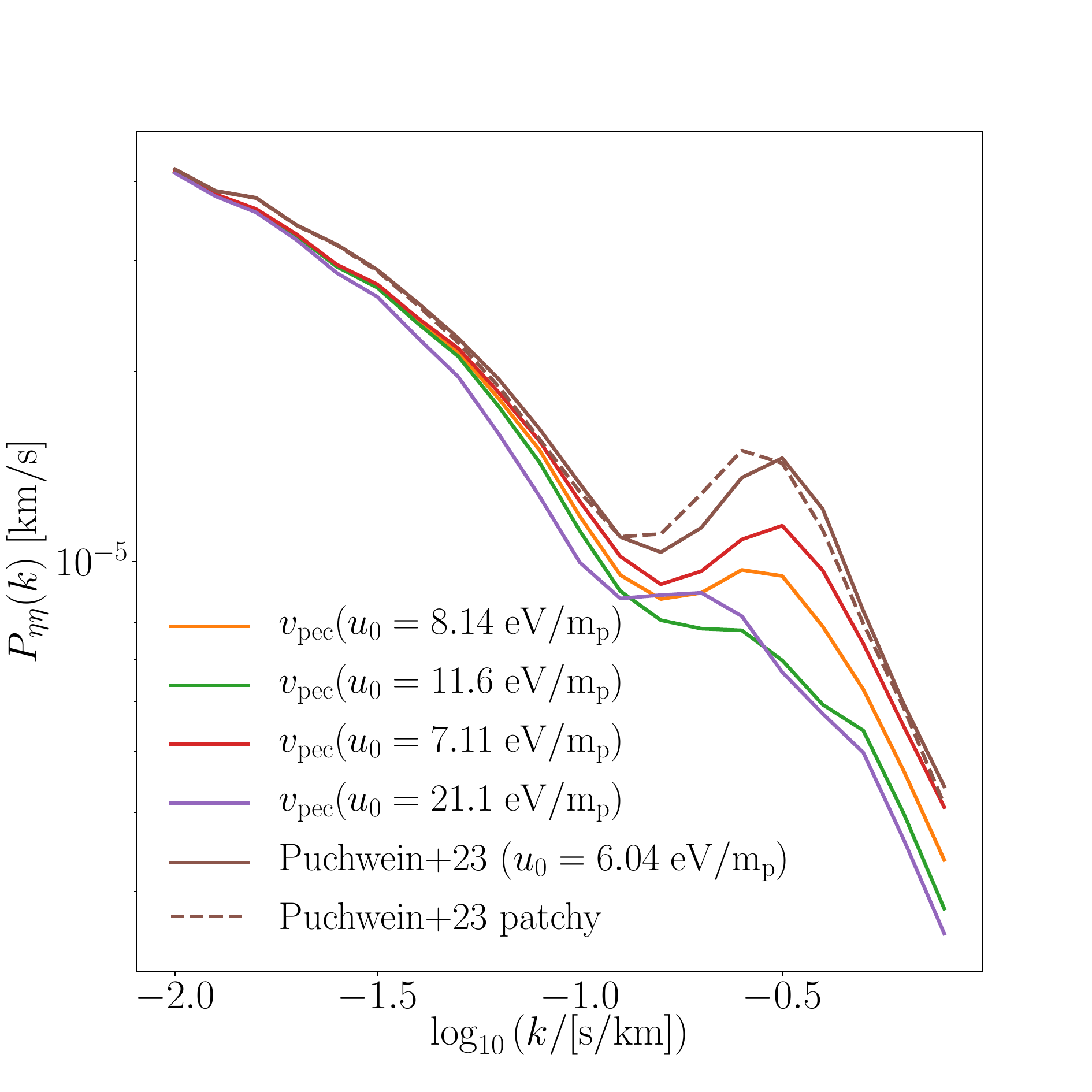}
\caption{ Effect of peculiar velocities on the small scale ($k>0.1\;\skm$) suppression of power. {\it Left:} The ratio of 1D flux power spectra of very early ($z_{\rm rei,end}=7.5$) and reference ($z_{\rm rei,end} = 6.0$) reionization models (solid blue). The dashed lines show the effect of replacing the peculiar velocity fields of both simulations simultaneously. The dashed blue line shows the effect where no peculiar velocities are included. The coloured lines show the effect of using peculiar velocity fields corresponding to different thermal histories -- in particular different heat injection values. All models show a relative increase of small-scale power ratio compared to the model with no peculiar velocities. The strength of this relative increase correlates with heat injected during reionization, with $v_{\rm pec}$ coming from a late reionization run (low heat injection; red dashed line) giving the strongest signal. {\it Right:} The relative increase in the small-scale structure of the 1D flux power spectrum is related to small-scale structure in the peculiar velocity field. A feature is present in the power spectrum of the peculiar velocity gradient ($\eta = \nabla v_{\rm pec}$), where the peak shifts from $k\sim 0.15\;\skm$ for early reionization models (with higher heat injection) to $k\sim 0.30\;\skm$ for late reionization models (with lower heat injection). Models with cumulative injected heat of $u_0=8.14, 11.6, 7.11, 21.1\;\mathrm{eV/m_p}$ correspond to the L20-ref, L20-very early, L20-late, L20-very early-hot models respectively. Similarly, the models from \citep{Puchwein2023} with $u_0=6.04\;\mathrm{eV/m_p}$ correspond to their late reionization model ($z_{\rm rei,end}=5.3$).} \label{fig:vpec_small_scales}
\end{figure*}

The enhancement of the small-scale power in the models is associated with the enhancement of the small-scale structure in the peculiar velocities. Fig.~\ref{fig:vpec_small_scales} (left) shows the relative effect of the peculiar velocity field on the flux power spectrum ratios. The models of early ($z_{\rm rei}=7.5$) and reference ($z_{\rm rei}=6.0$) reionization (blue solid line) show a relative enhancement of power compared to the ratio of the two flux power spectra when the effects of peculiar velocities are not included in the calculation of the optical depth. This effect of setting $v_{\rm pec}=0$ has also been seen in \citep{Villasenor2022b}. Fig.~\ref{fig:vpec_small_scales} further illustrates that the amplitude of this feature at $k>0.1\;\skm$ is sensitive to the amplitude of the peculiar velocity field changing with the amount of pressure smoothing. Furthermore, the feature in the flux power can be associated with the emerging feature in the 1D power spectra of the peculiar velocities (Fig.~\ref{fig:vpec_small_scales}; right), with the strength of the feature exhibiting a positive correlation between its amplitude and the cumulative injected heat. That the feature is stronger for later ending reionization, and weaker for earlier reionization, suggests that this behaviour is due to the hydrodynamic response of the gas to the photo-heating.

In terms of the constraints on the WDM particle mass, this suggests that a certain caution has to be exercised when pushing the models to $k>0.1\;\skm$. While the peculiar velocity feature might be related and correlated with the existing thermal parameters, it is not a-priori obvious that this new scale in the model is properly covered within the range of exisiting simulations, and therefore not properly marginalized over.

\subsection{Thermal dependence of the mass resolution}
\label{sec:thermal_mass_resolution}

The results of Sec.~\ref{sec:pecvel_effects} show that a small-scale peculiar velocity structure can modify the amount of small-scale flux power. However, this is also the regime where the mass resolution ($R_s$) of our simulations has the biggest effect.

\begin{figure*}
    \begin{center}
        \includegraphics[scale=0.55]{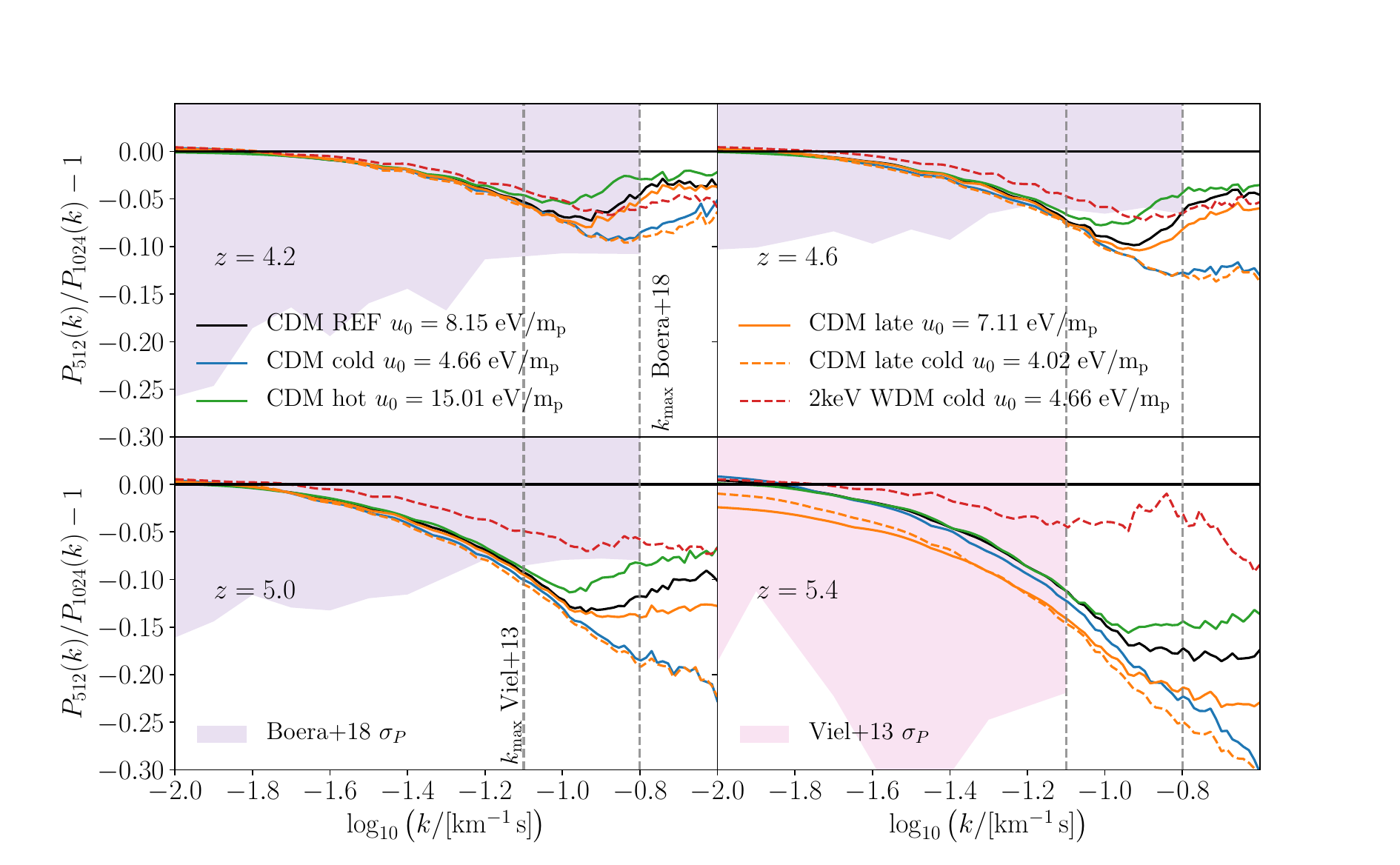}
    \end{center}
\caption{The effect of mass resolution in the simulations, shown as a flux power spectrum decrement as a function of wavenumber for different models with varying the thermal history and WDM particle mass. The more heat is injected into the IGM, the more the gas is smoothed due to pressure effects. Larger pressure smoothing results in less structure in the flux power spectrum at small scales, and depends less on the mass resolution. The same happens if the matter density is smoother due to the presence of free-streaming, which results in a smaller required mass resolution correction.}    \label{fig:mass_resolution_thermal}
\end{figure*}

The mass resolution correction of the simulations should depend on the thermal history in this high-$k$ regime of the model, i.e. $R_s=R_s(u_0)$. This is perhaps not surprising -- the mass resolution corrections essentially describe how much small-scale structure is missing in the (power spectrum) statistics as a result of not resolving the structure at very small scales, and its non-linear coupling to larger scales. If the field in configuration space is smoothed out due to physical effects -- such as higher pressure smoothing or larger free-streaming scale -- the amount of missing small-scale flux power will also be smaller. To estimate this effect we repeated the resolution correction exercise for different models in our suite of simulations (R10-; see Table~\ref{table:summary_sims}). The results, in Fig.~\ref{fig:mass_resolution_thermal}, show that indeed the mass resolution correction exhibits a strong dependence on the thermal history at $k>0.1\;\skm$. In particular, late reionization models with less pressure smoothing (or lower cumulative injected heat) can show up to 5\% larger mass resolution corrections compared to the fiducial correction used in the analysis. This trend is more prominent at higher redshifts, and less important at $z \leq 4.2$. On the other hand, models with larger pressure smoothing scales require consistently smaller resolution corrections at small-scales, by up to 2\%. 

Similarly to the effect of the thermal history, the smoothing of the density field due to free-streaming also decreases the required mass resolution correction. As shown in Fig.~\ref{fig:mass_resolution_thermal} a 2 keV WDM model on average requires a 5\% lower mass resultion correction at $k\sim 0.1\;\skm$ at $z=5.0$. This effect is reduced at lower redshifts.

\begin{figure*}
\centering
    \includegraphics[width=0.45\textwidth]{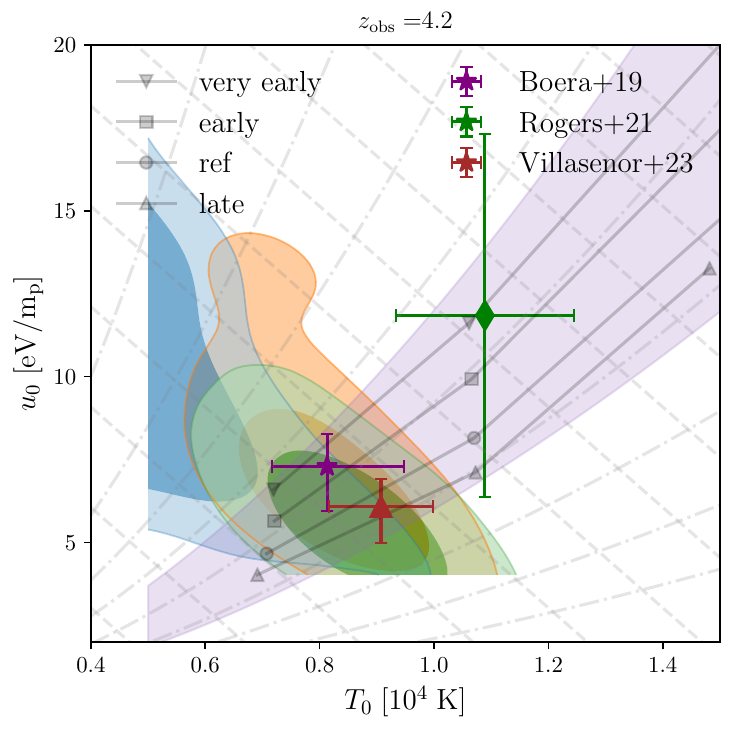}
    \includegraphics[width=0.45\textwidth]{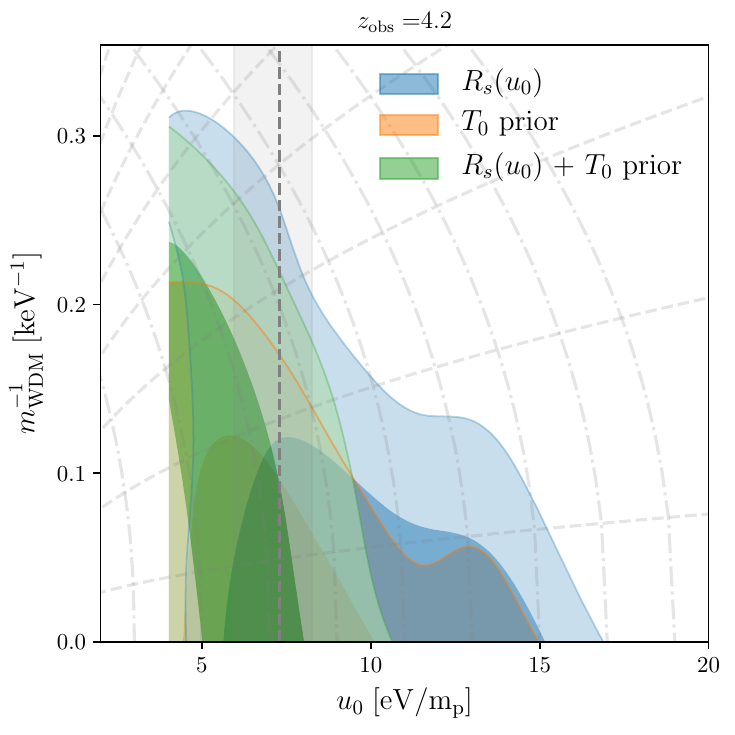}
\caption{Effect of using a mass resolution correction that depends on the IGM thermal history1 ($R_s(u_0)$). The two panels show 2D posterior distributions for redshift $z=4.2$ in the plane of temperature and heat injection (left) and warm dark matter mass and heat injection (right). In the thermal parameter space (left), the mass resolution correction adds more power for models with less pressure smoothing, which makes such models easier to fit the data. As a result the posterior still lies along the $u_0-T_0$ degeneracy axis, but the amplitude of this axis moves to lower $u_0$ values. Similarly, the degeneracy between free-streaming and pressure smoothing (right) opens up along the degeneracy axis, because the models with low $u_0$ require larger mass resolution corrections that increase the power, even for lower warm dark matter masses.
} \label{fig:effect_thermal_resolution}
\end{figure*}

These results imply that applying a mass resolution corrections that depends on the thermal history widens the range of $P_{\rm 1D}$ at $k>0.1\;\skm$ for models within a given section of the parameter space, ultimately resulting in higher sensitivity to thermal parameters and lower sensitivity to free-streaming. On the other hand, the mass resolution correction that depends on $m_{\rm WDM}$ shows stronger sensitivity of the $P_{\rm 1D}$ at $k>0.1\;\skm$, which leads to stronger constraints on the lower bound WDM mass. Current bounds on the WDM mass lie in the range of $\sim4-6\;\mathrm{keV}$, however, and the effect of mass resolution dependence on WDM free-streaming is severely reduced. Thus most of the effect of the mass resolution that depends on thermal history and WDM mass comes from the thermal history dependence.

The results of applying free-streaming and thermal history dependent mass resolution correction are shown in Fig.~\ref{fig:effect_thermal_resolution}. Compared to the analysis without any thermal priors shown in Fig.~\ref{fig:effect_thermal_priors}, the new mass resolution correction does not shift the posterior in the thermal parameters, suggesting that the effect of peculiar velocities is not completely explained by accounting for the thermal dependence in the mass resolution. On the other hand, the WDM constraints are weakened, roughly to the same level as when a sensible thermal prior is applied to the model, resulting in $m_{\rm WDM} > 4.44\;\mathrm{keV}$ at $2\sigma$. Low WDM masses of $3.19\;\mathrm{keV}$ and $2.78\;\mathrm{keV}$ are excluded at $3\sigma$ and $5\sigma$ respectively. Applying both the physical thermal prior ($T_0$ prior) and the new mass resolution correction leads to $m_{\rm WDM} > 4.24\;\mathrm{keV}$ at $2\sigma$.

\subsection{Patchy reionization}
\label{sec:patchy_reionization}

The original Sherwood suite of simulations used in this study evolves the reionization homogenously throughout the simulated volume. In reality the Universe reionizes in a more complex, inhomogenous manner, where local ionized bubbles first appear around the sources of ionizing photons \citep{Puchwein2023}. Observations of the \lya{} forest at higher redshifts can thus still be affected by relic fluctuations of the reionization persisting for a time after most of the Universe has been reionized. This topic has been a focus of several studies over the years \citep{Keating2019,Onorbe2019,Wu2021,Molaro2022,Cain2021,Molaro2023}. The main effect of the patchy nature of reionization on the 1D flux power spectrum of the \lya{} forest has been found to be an enhancement of power on large scales, that traces the fluctuations in the temperature and ionized fraction of hydrogen gas. The conclusions of recent works \citep{Molaro2022,Molaro2023} suggest that the enhancement of power appears at larger scales ($k<5\times 10^{-3}\;\skm$) than those observed in \citep{Boera2019}, i.e., the flux power spectrum measurements used in this study.

While the large scale effect of ionization fluctuations and their effect on the \lya{} forest has seen a certain agreement between different methods and simulations, the same is not true for the effect of inhomogenous reionization on small scales. Spatial fluctuations of the photo-ionization rate result in spatial fluctuations of the temperature density relation. Regions of the IGM that are ionized later heat up later as well, while regions that reionized and heated up earlier had time to cool down, due primarily to the expansion of the Universe and inverse Compton scattering \citep{Mcquinn15,Keating2019,Onorbe2019}. As a result regions ionizing later would exhibit stronger suppression of the flux power spectrum due to thermal Doppler broadening, than IGM regions that have ionized long ago. As pointed out by \citep{Wu2021,Puchwein2023} a competing effect to the thermal fluctuations, is that the IGM regions that ionized earlier had more time to hydrodynamically respond to the injected heat, resulting in a larger pressure smoothing scale. More pressure smoothing also reduces the small-scale power. It has been suggested that these two effects might largely cancel each other out, leaving small-scale power unchanged compared to homogenous reionization models.

In this study we make use of the tabulated correction to the 1D \lya{} flux power spectrum from \citep{Molaro2022}, that is based on the Sherwood Relics simulation suite \citep{Puchwein2023}. The effect of a patchy reionization correction in that study results in a $\sim10$\% suppression of \lya{} flux power at $k>0.1\;\skm$. The amplitude and shape of the suppression are largely independent of the thermal history models used in that study. Aside from the temperature fluctuations, \citep{Molaro2022} found that the dominant effect of the small-scale suppression was due to the effect of the peculiar velocity field.

\begin{figure*}
\centering
    \includegraphics[width=0.45\textwidth]{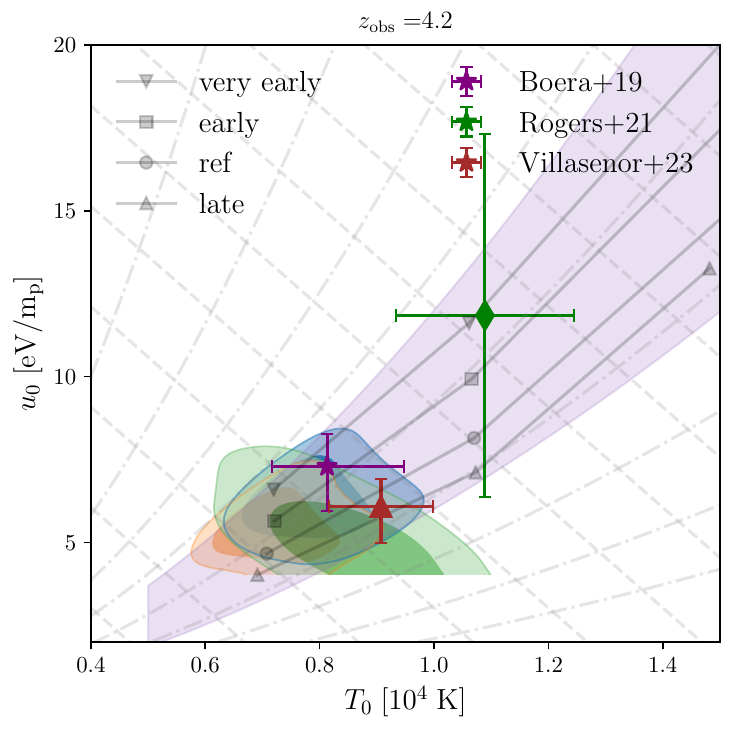}
    \includegraphics[width=0.45\textwidth]{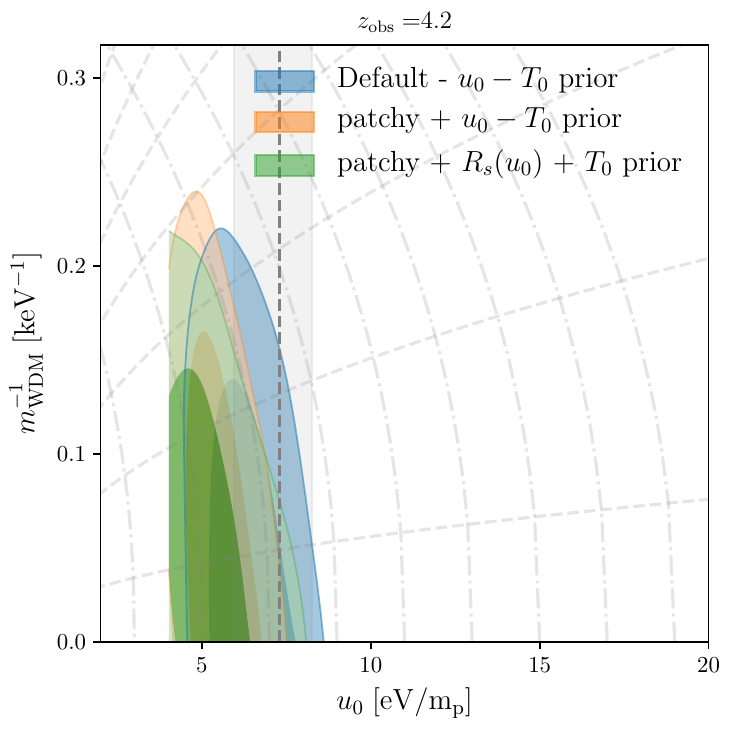}
\caption{Effect of including a correction due to inhomogenous reionization. The two panels show 2D posterior distributions for redshift $z=4.2$ in the plane of temperature and heat injection (left) and warm dark matter mass and heat injection (right). The effect on the analysis mostly comes from the small scale suppression of power due to the peculiar velocity structure of the gas, rather than the large scale enhancement of power due to photo-ionization and temperature fluctuations. Different thermal history models have an almost identical suppression of power due to inhomogenous reionization compared to the equivalent homogenous reionization model. The net effect is a systematic shift along the (positive) $u_0-T_0$ degeneracy axis in the thermal parameter space (left), in the direction of lower temperature and lower pressure smoothing. Similarly, the degeneracy between free-streaming and pressure smoothing (right) opens up along the degeneracy axis, because the models with lower thermal and pressure smoothing leave more freedom for WDM suppression to accommodate the data.}
\label{fig:effect_patchy}
\end{figure*}

The results of this analysis are presented in Fig.~\ref{fig:effect_patchy}. Since patchy reionization suppresses the small-scale power, one could expect that lower values of WDM masses might be even further excluded by the data. However the small-scale suppression induced by inhomogenous reionization affects {\it all models equally}, including the models with different thermal and pressure broadening. The main effect on the \lya{} data analysis is to move the peak of the $T_0$ posterior to lower values. The reason for this is as follows: the higher the $T_0$ value the stronger the suppression due to thermal broadening. The flux power spectrum models for low $T_0$ values that on their own do not exhibit enough suppression to explain the data, now achieve enough suppression through patchy reionization correction. Therefore the first conclusion is that lower $T_0$ models that were excluded before now fit the data, and the posterior of the $T_0$ parameter expands towards lower $T_0$ values. On the other hand, the models with high $T_0$ values would now show too strong of a suppression, and models that fit the data without the correction due to patchy reionization are now in tension with the data. The posterior of $T_0$ therefore shrinks for high $T_0$ values. Due to the prior in the $u_0-T_0$ plane, shifting the $T_0$ posterior to lower values also shifts the $u_0$ posterior to lower values at each redshift, resulting in data preferring less pressure smoothing. Since both the thermal and pressure smoothing effects are reduced, the posterior of the WDM mass expands to compensate for the fact that somewhat lower WDM mass models are now no longer in tension with the data.

While the specific result shown in Fig.~\ref{fig:effect_patchy} depends on the choice of thermal priors, the main conclusion would remain the same even in the light of less stringent priors. As the $T_0$ posterior systematically shifts to lower values, the $u_0-T_0$ anti-correlation direction is preserved as it depends on the fact that both parameters increase the small-scale suppression. The resulting posterior in a scenario with wider thermal priors would therefore only extend further along the $u_0-T_0$ anti-correlation direction, but still resulting in reduced sensitivity to $m_{\rm WDM}$.

From Fig.~\ref{fig:effect_patchy} we also observe that a $\sim10$\% suppression of power in all the models results in only $\sim0.5\sigma$ shift of the posterior in the $u_0-T_0$ plane, along the positive degeneracy axis (shift between the blue and orange countours). The constraints on the WDM mass are thus slightly weaker, with $m_{\rm WDM} > 5.10\;\mathrm{keV}$ at $(2\sigma)$.  However, as was highlighted in \citep{Molaro2022}, the 10\% small-scale suppression is mainly driven by the peculiar velocity field differences between the inhomogenous and homogenous reionization models. As we have shown in previous sections, the exact nature of the peculiar velocity structure on small-scales has implications for the WDM mass inference, and can affect both the mass resolution correction of the simulations as well as parametrisation of the thermal history on the smallest scales probed by the \lya{} forest ($\sim50-100\;\mathrm{ckpc/h}$). While the nature of the peculiar velocity field structure requires further study, it is reassuring that the effect on the WDM constraints is small ($\sim 10$\%). 

Combining the corrections due to inhomogenous reionization and the thermal history dependence of the mass resolution ($R_s(u_0)$), with the thermal priors coming from independent $T_0(z)$ observations ($T_0$ prior) we get a combined constraint on the WDM particle mass of $>5.9\;\mathrm{keV}$ (95\% C.L.). Even though individually both the patchy reionization and the $R_s(u_0)$ correction reduce the WDM constraining power, together with the $T_0$ prior they are pushed in the parameter space of higher $T_0$ values and a lower pressure smoothing scale (or late reionization), which leaves little room for additional suppression due to WDM free-streaming. While these constraints are the strongest presented in this paper, they rely on our first attempt at both a patchy reionization and $R_s(u_0)$ corrections. With their impact on the WDM particle mass, these results provide additional incentive to improve on the modelling of the small-scale thermal history in the inhomogenous reionization models.

\subsection{Instrumental effects}

Several instrumental and observational effects can potentially systematically alter the small-scale flux power spectrum: mis-estimation of the observed flux noise, contamination due to metal lines or the instrument resolution. Of the three, the instrument resolution has been one of the more studied effects, as it has a large impact on large \lya{} surveys that observe spectra at lower spectral resolution \citep{Palanque13,Irsic2017a,Karacayli2022,Wilson2022}. For a typical line-spread function shape the correction of the instrument resolution on the flux power spectrum is well described by a Gaussian kernel $P_F \rightarrow P_F / W_k^2 = P_F \exp{k^2 \sigma_R^2}$, with the Gaussian width of the resolution $\sigma_R=\mathrm{FWHM}_{R}/(2\sqrt{2\ln{2}})$ given as function of the FWHM resolution element ($\mathrm{FWHM}_{R} = c/R$) or resolving power ($R$). For the scales of $k\gtrsim \sigma_R^{-1}$ the correction due to resolution becomes $\sim 1$, dominating the total signal in the instrument. While of significant concern for lower resolution instruments such as X-Shooter ($R\sim8,000$), the resolving power of Keck/HIRES and VLT/UVES is high enough ($R\sim50,000-80,000$) to not play a major role in flux power spectrum measurements for scale cuts $k<0.1\;\skm$ \citep{Viel13wdm,Walther2018,Boera2019}. Indeed assuming the fiducial value of the resolution $\mathrm{FWHM}_{R} = 6\;\mathrm{km/s}$ ($R\sim50,000$) of the data \citep{Boera2019}, this translates into $\sigma_R = 2.55\;\mathrm{km/s}$. In order to improve the fit to the data at small scales, the resolution width would have to be overestimated by 30-40\%. Typically the resolution is estimated to $\sim10 \%$, and a factor three to four seems unlikely to be an explanation for excess small scale power.

Similarly, the contribution from contaminating metal absorption in the \lya{} forest has been studied in both low-resolution \citep{Palanque13,Irsic14} and high resolution data \citep{Irsic2017a,Boera2019,Wilson2022}. The contamination can be split into two main groups: (a) metals that have a rest-frame wavelength transition close to the \lya{} line (e.g. SiIII) \citep{McDonald05,Palanque13}; and (b) metals situated at a lower redshift and associated with either IGM or circumgalactic medium (CGM) contributions \citep{McDonald05,Tie2022}. The first group (a), imprints an oscillatory feature on the flux power spectrum. The frequency of this feature increases with scale, and is typically averaged over many periods in measurements of the high-$k$ flux power spectra, leaving distinct features observable only at low k, $k<0.01\;\skm$. The second group (b) is important at all redshifts and scales, and due to large differences in redshift can be subtracted statistically by measuring the flux power spectrum on the red side of the \lya{} emission line. The metal flux power spectra are typically dominated by CIV and SiIV doublets \citep{Karacayli2023}, and are smaller than the \lya{} flux power spectrum by one or two orders of magnitude. The amplitude of the metal power spectrum would need to be larger by a factor of 5-10, in order to have an impact on \lya{} flux power spectrum parameter estimation. While some studies suggest that the redside metal power spectrum captures only about half of the contaminated metal content in the \lya{} forest \citep{Day2019}, it is difficult to argue on observational grounds that the small-scale enhancement of the small-scale power spectrum due to metals could significantly affect our analysis.

The flux noise estimation has received somewhat less attention as a source of systematic uncertainty in high resolution and high signal-to-noise quasar spectra. It plays a crucial role in the low signal-to-noise spectra of large surveys (e.g. \citep{Palanque13}). The flux power spectrum of the noise is 2-3 orders of magnitude lower than the \lya{} forest flux power spectrum signal in medium ($S/N>20$) and high ($S/N>40$) quality data (e.g. \citep{Irsic2017a}). The signal decays exponentially towards higher wave numbers, suggesting that the noise flux power quickly becomes a bigger contribution to the signal as the analysis is pushed towards higher k.

\begin{figure*}
    \centering
    \includegraphics[width=1.0\textwidth]{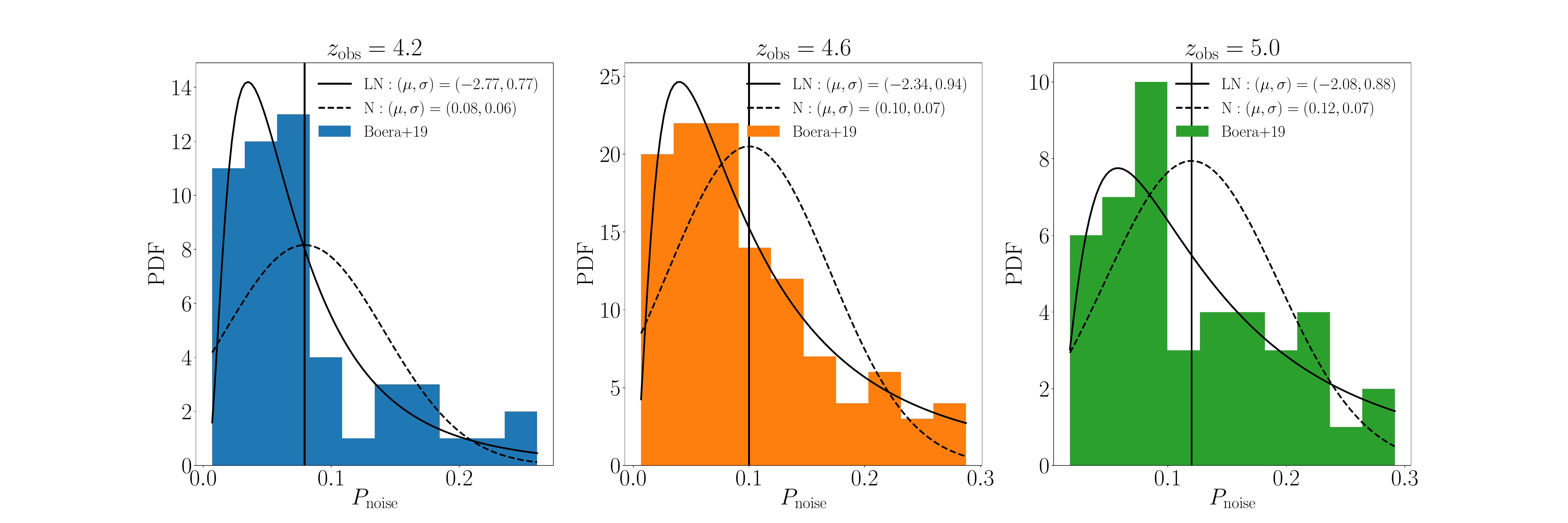}
    \caption{The probability distribution of the measured noise power spectra from \citep{Boera2019} in each of the redshift bins. The measured distribution is reasonably well approximated by a log-normal distribution at each redshift, shown as a solid black line. For comparison we also show a normal distribution with mean and variance computed from the first two moments of the measured distribution. The distributions are fairly broad, however the inclusion or removal of tails beyond the range of measured $P_{\rm noise}$ has a negligible effect.} \label{fig:noise_pdf}
\end{figure*}

The analysis of \citep{Boera2019} estimated the noise power on a per quasar sightline basis in $20\;\Mpch$ sections. This was achieved by measuring the raw or total flux power spectrum in each section of the \lya{} forest and estimating the asymptote level at high k. This method relies on the assumption that the noise power is white -- an assumption that is largely validated in other studies (e.g. \citep{Palanque13,Irsic2017a,Yeche2017}) --  and that it dominates at high wavenumbers. The method contends with several challenges, from a noisy estimation of the measured noise power spectrum in individual $20\;\mathrm{cMpc/h}$ sections, to the fact that the asymptote levels at high k will also include the metal contamination, as well as the very signal that one wishes to measure. 

A careful analysis of uncertainty propagation is warranted, especially for a signal dominated by the highest wavenumbers such as is the case in this WDM study. Fig.~\ref{fig:noise_pdf} shows the probability distribution of the noise power estimates from the individual $20\;\mathrm{cMpc/h}$ sightline sections, in each of the three redshift bins. The vertical black lines indicate the effective average $P_{\rm noise}$ assumed in the analysis of \citep{Boera2019}. This is simply a result of averaging the difference of raw and noise power per section over all the sightlines in a given redshift bin. As the average was not weighted by the signal-to-noise, the estimated average $P_{\rm noise}$ is simply the mean of the distribution. One immediate conclusion of Fig.~\ref{fig:noise_pdf} is that the distribution of the noise power is not Gaussian around the mean, with the bulk of the distribution typically peaking at lower than average $P_{\rm noise}$ values. The distributions at each redshift are also relatively broad. The mean of the distributions, $\langle P_{\rm noise} \rangle$ are 0.08, 0.1 and 0.12 for $z=4.2$, $4.6$ and $5.0$ respectively. This corresponds to roughly 5\% of the total power at the highest wavenumber in the data. We approximate the $P_{\rm noise}$ distributions with a log-normal model with the min/max range of the measured values (solid black lines in each of the redshift bins in Fig.~\ref{fig:noise_pdf}).

The noise power spectrum distribution in Fig.~\ref{fig:noise_pdf} is dominated primarily by the distribution of signal-to-noise in the data, as well as the mean transmission variations among the $20\Mpch$ segments of the \lya{} forest. This has been verified in mock data with pathlength and redshift ranges of observed quasars reported in \citep{Boera2019}. The methodology of estimating the noise power asymptote within each $20\Mpch$ segment is noisy it is therefore unlikely that the uncertainty on the noise power estimation exceeds the width of the distributions in Fig.~\ref{fig:noise_pdf}. As such we use the distribution of the noise power as a conservative prior.

\begin{figure*}
\centering
    \includegraphics[width=0.45\textwidth]{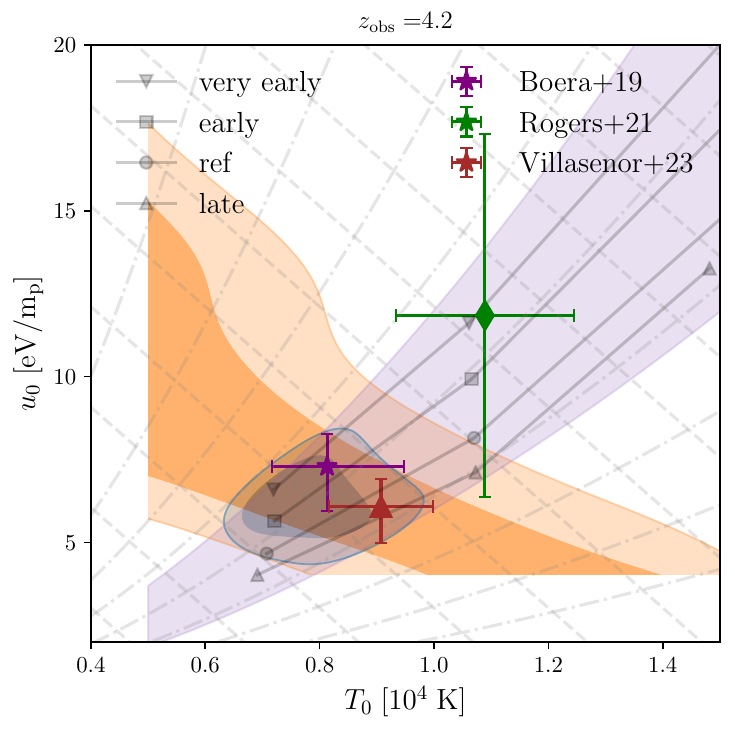}
    \includegraphics[width=0.45\textwidth]{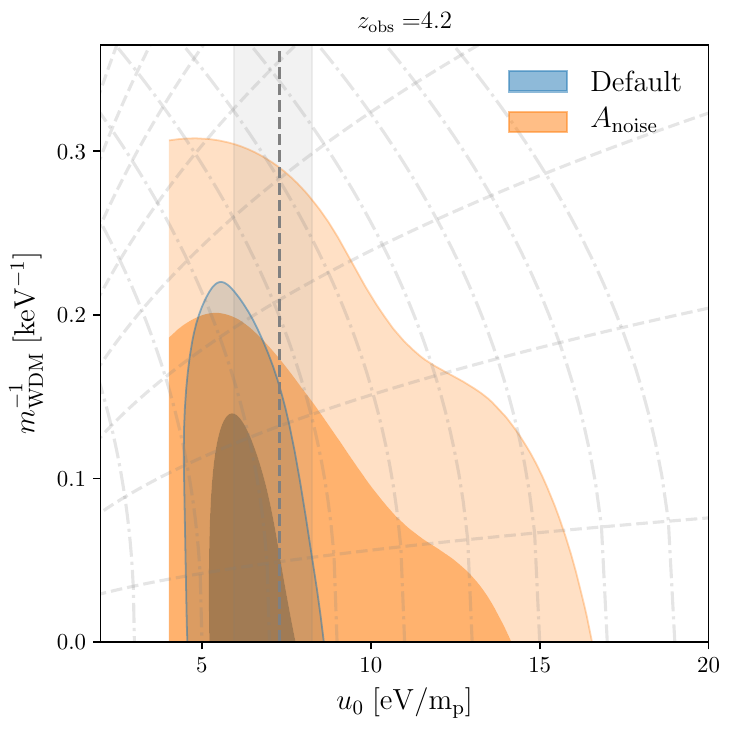}
\caption{Effect of marginalizing over the noise uncertainty distribution. The two panels show 2D posterior distributions for redshift $z=4.2$ in the plane of temperature and heat injection (left) and warm dark matter mass and heat injection (right). As the noise affects the amount of small-scale power it effectively removes the information from those scales, which leads to poorer constraints on thermal parameter as well as warm dark matter mass. The noise distribution is marginalized over the measured distribution from \citep{Boera2019}. However, the results remain largely unchanged if the shape of the distribution was changed to a normal distribution with a standard deviation of 10\% of the measured power. This can also be included at the level of the covariance matrix. }
\label{fig:effect_noise}
\end{figure*}

In order to asses the potential impact of noise mis-estimation in the data, we add a constant term $A_{\rm noise}(z)$ to the model of the 1D flux power spectrum. This term is scale independent, but is modelled separately for each redshift bin. Since the mean of the noise flux power spectra distributions were already subtracted from the data, this constant term measures the deviation of the noise flux power from this mean value. In the data analysis step, the noise is subtracted from the raw power before the resolution correction of the instrument is deconvolved. The theoretical 1D flux power spectrum model is modified as follows:
\begin{equation}
    P_{F,1D}^{\rm tot}(k,z) = P_{F,1D}^{\rm Ly\alpha}(k,z) + A_{\rm noise}(z) \frac{\langle P_{\rm noise} \rangle(z)}{W^2(k)},
\end{equation}
where $P_{F,1D}^{\rm Ly\alpha}$ is the \lya{} flux power spectrum as given by the emulator, $\langle P_{\rm noise} \rangle(z)$ are the means of the $P_{\rm noise}$ distribution in each of the redshift bins, and $W^2(k)$ is the instrumental correction due to resolution and pixel size (Following \citep{Boera2019} we use a pixel size of $2.5\;\mathrm{km/s}$ and a FWHM resolution of $6.0\;\mathrm{km/s}$ using top-hat and Gaussian kernels for the two corrections, respectively.).

Fig.~\ref{fig:effect_noise} shows the results of the analysis where three $A_{\rm noise}$ parameters were added to the theoretical model (one for each redshift bin), and the parameters' priors were assumed to be given by the approximate log-normal model of the $P_{\rm noise}$ distribution. The resulting WDM mass constraint is slightly weakened, and the lower WDM mass bound is $m_{\rm WDM} > 3.91\;\mathrm{keV}$. The thermal constraints are significantly degraded along the $u_0-T_0$ degeneracy axis. This is because at every individual redshift, the noise parameter $A_{\rm noise}$ strongly correlates (anti-correlates) with the IGM temperature (cumulative heat injection), whereas the correlation with the WDM mass parameter is weaker. The marginalized mean of the posteriors (and their best-fit) values of $A_{\rm noise}$ are $0.74_{-0.49}^{+0.49}$ ($0.24$), $1.12_{-0.29}^{+0.49}$ ($1.73$), $0.87_{-0.15}^{+0.56}$ ($1.38$). The data prefers values of $A_{\rm noise} > 0$, implying noise was underestimated. The best-fit values also show a slight increase with redshift, suggesting that the effect was larger for higher-redshift quasar spectra. The typical values of $A_{\rm noise}$ are of the order of unity, suggesting that the noise subtraction of the data performed by \citep{Boera2019} may be incomplete. The sensitivity of thermal parameters to this relatively small noise contribution is quite large, possibly implying that measurements of the IGM temperature and reionization are very sensitive to noise subtraction in the data. There is also sensitivity of the WDM mass constraints to this effect, although somewhat reduced compared to the thermal parameters.

The sampling of the $P_{\rm noise}$ distribution is relatively sparse, measured in $20\;\Mpch$ sections in only 15 quasar sightlines in each redhift bin. This marks a significant improvement on previous measurements, but is nonetheless sensitive to sample variance. To understand the sensitivity of the conclusions of this analysis step we modify the prior choice to be a Gaussian distribution, with the mean and the standard deviation estimated from the average and variance of the samples in each redshift bin. Since the posteriors of $A_{\rm noise}$ for the highest two redshifts are dominated by the upper limit on the prior range, we further allow this Gaussian prior to have no min/max limits other than the physical requirement that noise is larger or equal to zero $P_{\rm noise} \geq 0$. This allows for a tail of the $A_{\rm noise}$ distribution to arbitrarily large values. The mean of the posterior distributions of $A_{\rm noise}$ parameters however do not move significantly. The main difference is the tail of the posteriors towards higher $A_{\rm values}$. Note that the WDM constraint changes less than $1$\%.

If the signal at the high-$k$ end of \citep{Boera2019} data is indeed due to under-subtracted noise power, then the situation should improve with better and more data. If the flux uncertainties are dominated by the read-out noise component, then increasing the signal-to-noise ($S/N$) ratio of individual quasar sightlines quickly reduces the level of noise power ($P_{\rm noise} \propto (S/N)^{-2}$) \citep{McQuinn2011}. Future studies should thus suffer less from the impact of instrumental effects on the flux power spectrum measurements, allowing for exploration of data to high $k_{\rm max}$.

\begin{figure*}
\centering
    \includegraphics[width=0.45\textwidth]{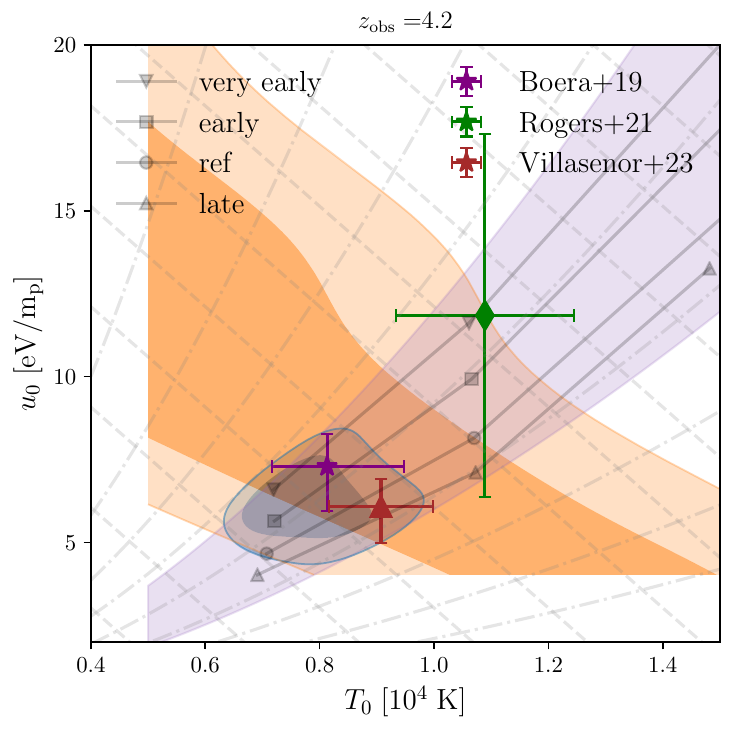}
    \includegraphics[width=0.45\textwidth]{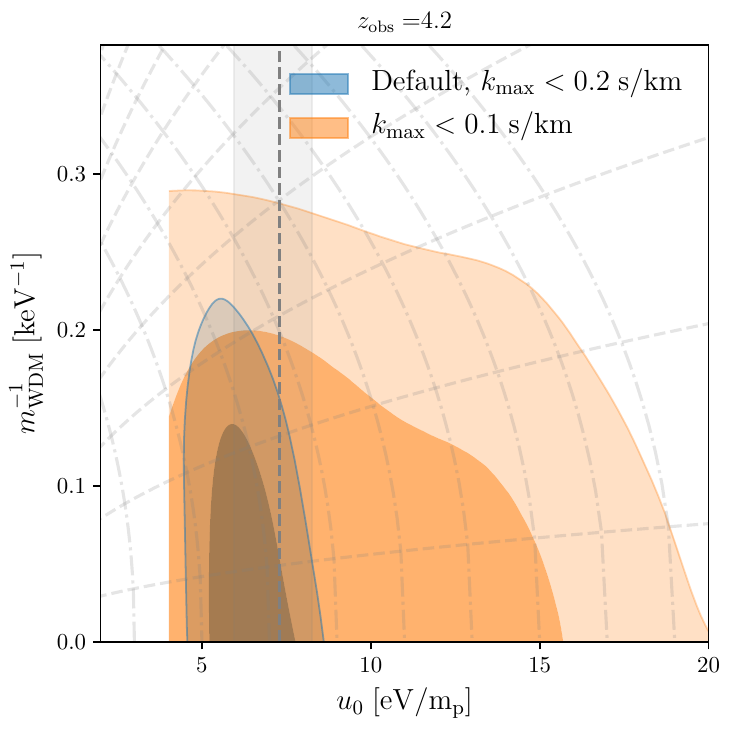}
\caption{Effect of varying $k_{\rm max}$ of the analysis. The two panels show 2D posterior distributions for redshift $z=4.2$ in the plane of temperature and heat injection (left) and warm dark matter mass and heat injection (right). In the thermal parameter space (left) limiting the analysis to the value of $k_{\rm max} < 0.1\;\skm$ -- similar to previous analyses using Ly$\alpha$ forest data -- has a similar effect to marginalizing over noise or thermal dependence of the resolution correction. The posterior stretches in the degeneracy direction of $u_0-T_0$. Similarly the warm dark matter mass constraints relax as more thermal support can accommodate the data.}
\label{fig:effect_kmax}
\end{figure*}

\subsection{Small-scale data cuts}

In order to facilite a more direct comparison between the new analysis using \citep{Boera2019} data, and previous analyses using high redshift HIRES/MIKE data \citep{Viel13wdm,Irsic2017b} a consistency check can be performed by limiting the new analysis to the same scale cuts ($k<0.1 \;\skm$). We further compare such an analysis to \citep{Irsic2017b,Viel13wdm} that used $k_{\rm max} = 0.088\;\skm$, and a similar redshift range. The previous analysis extended to $z=5.4$, however the flux power spectrum uncertainty at this highest redshift was considerably larger, and most of the constraining power came from the $z=4.2,4.6,5.0$ redshift bins, which are also the ones used in this study. Furthermore, we limit the comparison to the thermal history priors where $T_0$ is varied independently in each redshift bin. In \citep{Irsic2017b} the resulting lower bound on the WDM mass was $\sim 2.1\;\mathrm{keV}$ ($2\sigma$) (MIKE/HIRES Iršič+17 + wide thermal prior (Fig.~\ref{fig:wdm_constraints}); \citep{Irsic2017b}). 

A similar test was performed in \citep{Villasenor2022b}, where the reported value on the WDM mass bound sits at $3.6\;\mathrm{keV}$. The same scale cuts were used, using the quasar spectra data of \citep{Boera2019}. However somewhat different thermal history priors were applied.

Fig.~\ref{fig:effect_kmax} shows the result of small scale data cuts in this analysis. Using the same scale cuts and treatment of the thermal history with the new data improves the constraint to $m_{\rm WDM} > 4.09\;\mathrm{keV}$ ($2\sigma$) ($k_{\rm max} < 0.1\;\skm$ - this work (Fig.~\ref{fig:wdm_constraints})). The right hand side panel of Fig.~\ref{fig:effect_kmax} illustrates that imposing conservative scale cuts reduces the sensitivity to $m_{\rm WDM}^{-1}$ and pressure smoothing scale as probed by the injected heat $u_0$. This reduced sensitivity to the pressure smoothing can be understood in the thermal parameter space (left panel of Fig.~\ref{fig:effect_kmax}) as expanding of the posterior along the $u_0-T_0$ degeneracy axis. The posterior in this parameter space also shifts by $\sim 0.2\sigma$ along the positive $u_0-T_0$ relation that exists in hydro-dynamical simulations. The shift, however, is small, and can at least in part be attributed to reaching the corner of the priors at low $u_0$ and low $T_0$ values. 

In the case of conservative scale cuts ($k_{\rm max} < 0.1\;\skm$) the best-fit improves over the fit to all the data, with the $\chi^2/\mathrm{d.o.f.} = 10.2/20$ (see Table~\ref{table:summary_mcmc}). The fit prefers slightly warmer temperatures of the IGM ($T_0(z=4.6) \sim 8,400\;\mathrm{K}$) and slightly higher value of cumulative injected heat. The posterior of the $u_0$ parameter is very wide, however, suggesting that with conservative scale cuts the data are not sensitive to this parameter anymore. This has been observed in previous analysis using older data sets that did not extend beyond $k\sim0.1\;\skm$.

\section{Discussion and Conclusions}
\label{sec:conclusions}

This study presents new constraints on the free-streaming of WDM using a simulation based likelihood and Bayesian analysis of the VLT/UVES and Keck/HIRES \lya{} forest flux power spectrum measurements of \citep{Boera2019}. The new constraints of our fiducial analysis on the mass of a thermal relic WDM particle mass, $m_{\rm WDM} > 5.7\;\mathrm{keV}$, are the strongest to date. For the fixed shape of the WDM transfer function used in this study, the bound on the WDM particle mass translates into a wavenumber scale below which the matter power spectrum cannot drop by more than 5\%, $k_{0.05}=14.35\;\Mpch$.

Comparing to the previous high-redshift \lya{} forest data from HIRES/MIKE \citep{Viel13wdm}, the new data comprises of a larger number of quasar spectra in the range $4.2<z<5.0$, and is probing small scales up to a wavenumber of $k_{\rm max}=0.19\;\skm$ -- almost a factor of two improvement. Limiting the analysis to the same $k_{\rm max}$ cuts, and thermal state priors, as in previous HIRES/MIKE analyses we find the constraint to be $m_{\rm WDM} > 4.1\;\mathrm{keV}$ (this work), compared to $m_{\rm WDM} > 2.0\;\mathrm{keV}$ (e.g. \citep{Irsic2017b}). This factor two improvement on the bound on the WDM particle mass is consistent with the expected improvement of the statistical power of the high-redshift \lya{} forest data, and WDM mass sensitivity at $0.1\;\skm$ dominated by statistical uncertainty. This result is qualitatively similar to the recent analysis of \citep{Villasenor2022b} using the same scale cuts, and a different thermal state parametrisation. 

Recent studies of \citep{garzilli19,Villasenor2022b} have found a preference for non-zero $m_{\rm WDM}^{-1}$ in their default analysis, indicating a preference for a WDM cosmology. This warrants further study and rigorous tests on both the data and theory side. Our findings here lead us to suggest that one possibility is that the non-zero preference in the WDM parameter space is a result of the restricted variations in the thermal parameters. E.g., the analysis of \citep{garzilli19} assumes a thermal history with very low cumulative injected heat, and therefore a small amount of pressure smoothing. Limiting the amount of pressure smoothing can be of interest in specific applications, but in terms of a WDM particle mass constraint a thorough marginalization over the parameter space should be more robust. The analysis of \citep{Villasenor2022b} follow a similar simulation setup and parameter space variation as in this work, except for two main differences: (a) the variations in the redshift of hydrogen reionization were much narrower than explored in this work, and (b) parametrisation in the $z_{\rm rei}-T_0$ as opposed to $u_0-T_0$ plane only allowed for more restricted thermal histories.

The HIRES/UVES data of \citep{Boera2019} has also recently been used to provide constraints on a slightly different class of dark matter models -- ultra-light axion dark matter \citep{Rogers2021}. While the transfer functions of the two dark matter models are different enough that a direct comparison is non-trivial, the results of \citep{Rogers2021} suggest a strong bound on the thermal WDM mass, while at the same time recovering a hotter thermal history compared to both results in the literature \citep{Boera2019,Villasenor2022b} and the results of this study. A more thorough investigation is required but a major difference in the simulation setup of \citep{Rogers2021} is the initial condition generation with {\tt MP-Gadget} \citep{MPGadget} which uses glass initial conditions for the gas component. This introduces spurious small-scale power \citep{Khan2023}. 

Additionally, \citep{Rogers2021} uses a \lya{} spectral extraction code {\tt fake\_spectra} \citep{fakespectra} that uses a different optical depth assignment scheme that leads to additional enhancement of small-scale power in the 1D \lya{} forest flux power spectrum \footnote{Personal correspondence with S. Bird.}. These differences in the simulated flux power spectrum suggest further investigation is required for the comparison with the ultra-light axion constraints of \citep{Rogers2021}.

Further improvement in the WDM constraint comes from the smallest scales, $k>0.1\;\skm$. The sensitivity to the WDM mass is increased at smaller scales, resulting in potentially stronger constraining power. However, the regime of $k>0.1\;\skm$ is also more sensitive to observational and modelling systematics. In this study we have reviewed several aspects of the observational and instrumental systematics, of which the observational flux noise subtraction in the \lya{} forest flux power spectrum is potentially the most likely to affect the results. An average of a factor of two increase in the noise power (or 40\% increase in the level of noise), would on its own explain the small-scale signal observed, with no additional cosmological information beyond $k_{\rm max}>0.1\;\skm$. While this appears not very likely, it illustrates the need for improved treatment of the noise power at smallest scales in future \lya{} forest data analyses.

The thermal history priors have previously been identified as the dominant source of modelling systematics in the \lya{} forest flux power spectrum. In this study we revisited this, by exploring thermal priors motivated by different assumptions: a prior in the plane of IGM temperature and cumulative injected heat that envelopes physically motivated simulations consistent with the still rather weak constraints on the evolution of the neutral hydrogen fraction during the epoch of reionization, or a simple prior on the IGM temperature as interpolated from the measurements of the IGM temperature at $z<4.2$ and $z>5.0$. This was possible due to the improved range of simulations as well as post-processing techniques to expand on the number of models. While the posterior distributions are indeed sensitive to the choice of these priors, the WDM mass only changes by 2\%. This suggests that reasonable thermal priors lead to a stable and robust constraint on the WDM particle mass. We further point out that not imposing any thermal priors leads to a stronger and not weaker bound on the WDM particle mass.

With the data extending to $k>0.1\;\skm$, we have also identified a new source of modelling systematics -- the gas peculiar velocity field as modified by inhomogenous reionization. The effect of peculiar velocities is present, although different in amplitude, in both homogeneous and inhomogeneous models of reionization. The peculiar velocity field induces a knee in the flux power spectrum, that appears sensitive to the cumulative injected heat, indicating that the timing and process of reionization are an important factor in the peculiar velocity structure. This high-$k$ regime of the models is also sensitive to the numerical mass resolution of the simulations (at a level of up to 20\%). The peculiar velocity field structure is sensitive to the mass resolution at the level of as much as 50\%, resulting in thermal history dependent mass resolution corrections. All of these statements result in a similar effect on the WDM mass inference, weakening the constraint to $m_{\rm WDM} > 4.44\;(5.10)\;\mathrm{keV}$ at 95\% C.L. for including the mass resolution and inhomogeneous reionization respectively. Combining inhomogenous reionzation and mass resulution corrections together with observationally motivated thermal prior results in a WDM constraint that is not very different from our fiducial analysis, while at the same time preferring a slightly hotter IGM and reionization histories that end later. These statements are, however, somewhat model and simulation dependent, and indicate that further work into the origin and impact of the small-scale peculiar velocity structure is required. However, an important result for the particle astrophysics modelling is that WDM particle masses of $2.5\;\mathrm{keV}$ are ruled out at more than $5\sigma$, and $3\;\mathrm{keV}$ at more than $3\sigma$, for any of the analysis choices presented in this paper. In fact, the $3\;\mathrm{keV}$ is ruled out at $5\sigma$ for any of the reasonable choices of thermal priors (e.g. a $T_0$ prior).

We summarize the main conclusion points as follows:

\begin{itemize}
    \item Our fiducial analysis leads to improved WDM mass constraints from high-redshift quasar spectra of $m_{\rm WDM} > 5.7\;\mathrm{keV}$ at 95\% C.L.
    
    \item Using small scale data cuts, limiting the analysis to $k_{\rm max} < 0.1\;\skm$, results in a WDM constrain of $m_{\rm WDM} > 4.1\;\mathrm{keV}$ at 95\% C.L., a factor of two stronger constraint than previously published for the same choice of thermal priors and redshift range of the data \citep{Viel13wdm,Irsic2017b}.
    
    \item The 50\% higher WDM constraint coming from small-scales $k>0.1\;\skm$ has been explored with a variety of checks for instrumental systematics. We find that the flux noise may be underestimated by 40\% in the data, reducing the constraining power at $k>0.1\;\skm$. It should be possible to mitigate this in future surveys by a careful study of the instrumental noise, as well as by obtaining higher signal-to-noise spectra.
    
    \item The modelling uncertainties on the small-scale peculiar velocity structure can weaken the constraining power on the WDM mass by as much as $25$\%. The effect of thermal history and inhomogenous nature of reionization on the peculiar velocity fields of the baryonic gas is still poorly explored and needs further study.
\end{itemize}

The \lya{} forest data continue to push the frontier on astrophysical constraints on the CDM paradigm. The new constraints on the free-streaming of dark matter in this study both improve on exisiting constraints, and demonstrates that a larger number of quasar sightlines should translate into strong improvements on the WDM particle mass bound. As has been done in previous studies \citep{Viel13wdm,Irsic2017b,palanque20}, the high-redshift \lya{} forest data can be further combined with the low-redshift ($z<4.0$) flux power spectrum measurements in order to increase the redshift leverage arm, and further push the constraints on the free-streaming of dark matter. We leave such a study for future work.

\appendix
\renewcommand\thefigure{\thesection.\arabic{figure}}
\setcounter{figure}{0}

\bigskip

\acknowledgments 

The authors would like to thank Matthew McQuinn, Simeon Bird, Steven Gratton and Anson D'Aloisio for helpful conversations.

VI acknowledges support by the Kavli Foundation. MV is supported by the INFN PD51 INDARK grant. Support by ERC Advanced Grant 320596 ‘The Emergence of Structure During the Epoch of Reionization’ is gratefully acknowledged. MGH has been supported by STFC consolidated grant ST/N000927/1 and ST/S000623/1. JSB is supported by STFC consolidated grants ST/T000171/1 and ST/X000982/1. GB is supported by the National Science Foundation through grant AST-1751404. GK is partly supported by the Department of Atomic Energy (Government of India) research project with Project Identification Number RTI~4002, and by the Max Planck Society through a Max Planck Partner Group. For the purpose of open access, the author has applied a Creative Commons Attribution (CC BY) licence to any Author Accepted Manuscript version arising from this submission.

The simulations used in this work were performed using the Joliot Curie supercomputer at the Tré Grand Centre de Calcul (TGCC) and the Cambridge Service for Data Driven Discovery (CSD3), part of which is operated by the University of Cambridge Research Computing on behalf of the STFC DiRAC HPC Facility (www.dirac.ac.uk). We acknowledge the Partnership for Advanced Computing in Europe (PRACE) for awarding us time on Joliot Curie in the 16th call. The DiRAC component of CSD3 was funded by BEIS capital funding via STFC capital grants ST/P002307/1 and ST/R002452/1 and STFC operations grant ST/R00689X/1. This work also used the DiRAC@Durham facility managed by the Institute for Computational Cosmology on behalf of the STFC DiRAC HPC Facility. The equipment was funded by BEIS capital funding via STFC capital grants ST/P002293/1 and ST/R002371/1, Durham University and STFC operations grant ST/R000832/1. DiRAC is part of the National e-Infrastructure.

\bibliography{references}

\end{document}